\documentclass[11pt]{article}
\usepackage{amsmath,amsfonts,amsthm,amssymb}

\newcommand\version{August 31, 2006}

\newcommand\Z{{\mathbb Z}}
\newcommand\N{{\mathbb N}}
\newcommand\C{{\mathbb C}}
\newcommand\R{{\mathbb R}}
\newcommand\eps{\varepsilon}
\newcommand\half{\mbox{$\frac 12$}}
\newcommand\const{{\rm const.\, }}
\newcommand\Tr{{\rm Tr}}
\newcommand\tr{{\rm tr}}
\newcommand\Ff{{\cal F}}
\newcommand\Hh{{\cal H}}

\newcommand\vu{\nu}
\renewcommand\kappa\varkappa
\renewcommand\rho\varrho

\def\an#1{a_{#1}^{\phantom{\dagger}}}
\def\ad#1{a_{#1}^\dagger}
\newcommand\dH{{\mathbb H}}
\newcommand\dT{{\mathbb T}}
\newcommand\dV{{\mathbb V}}
\newcommand\dK{{\mathbb K}}
\newcommand\dW{{\mathbb W}}
\newcommand\dw{{w}}
\newcommand\dN\N
\newcommand\Bb{{\mathcal B}}
\newcommand\Aa{{\mathcal A}}

\newcommand\Ss{{\mathbb S}}

\def\Xint#1{\mathchoice
   {\XXint\displaystyle\textstyle{#1}}
   {\XXint\textstyle\scriptstyle{#1}}
   {\XXint\scriptstyle\scriptscriptstyle{#1}}
   {\XXint\scriptscriptstyle\scriptscriptstyle{#1}}
   \!\int}
\def\XXint#1#2#3{{\setbox0=\hbox{$#1{#2#3}{\int}$}
   \vcenter{\hbox{$#2#3$}}\kern-.5\wd0}}

\def\dashint{\Xint-}

\newtheorem{thm}{THEOREM}

\newtheorem{lem}{Lemma}

\begin{document}

\markboth{\scriptsize{\version}}{\scriptsize{\version}}
\title{\bf{Free Energy of a Dilute Bose Gas:\\ Lower Bound}}

\author{\vspace{5pt} Robert Seiringer
\\ \vspace{-2pt}\small{ Department of Physics, Jadwin Hall, Princeton
University,  }\\ \vspace{-2pt}\small P.O. Box 708, Princeton NJ
08544, USA. 
\\ {\small Email: \texttt  {rseiring@math.princeton.edu}} }

\date{\small \version}
\maketitle

\begin{abstract}
   A lower bound is derived on the free energy (per unit volume) of a
   homogeneous Bose gas at density $\rho$ and temperature $T$. In the
   dilute regime, i.e., when $a^3\rho \ll 1$, where $a$ denotes
   the scattering length of the pair-interaction potential, our bound
   differs to leading order from the expression for non-interacting
   particles by the term $4\pi a ( 2\rho^2 - [\rho-\rho_c]_+^2 )$.
   Here, $\rho_c(T)$ denotes the critical density for Bose-Einstein
   condensation (for the non-interacting gas), and $[\, \cdot \, ]_+ =
   \max\{ \, \cdot\, , 0\}$ denotes the positive part. Our bound is
   uniform in the temperature up to temperatures of the order of the
   critical temperature, i.e., $T \sim \rho^{2/3}$ or smaller. One of
   the key ingredients in the proof is the use of coherent states to
   extend the method introduced in \cite{RSjellium} for estimating
   correlations to temperatures below the critical one.
\end{abstract}

\renewcommand{\thefootnote}{${\,}$}
\footnotetext{Work partially supported by U.S. National Science
Foundation grant PHY-0353181 and by an Alfred P. Sloan Fellowship.}
\renewcommand{\thefootnote}{${\, }$}
\footnotetext{\copyright\,2006 by the author.
This paper may be reproduced, in its entirety, for non-commercial
purposes.}

\numberwithin{equation}{section}

\section{Introduction and Main Result}

The advance of experimental techniques for studying ultra-cold atomic
gases has triggered numerous investigations on the properties of
dilute quantum gases. From a mathematical point of view, several
rigorous results have been obtained over the last few years. (See
\cite{oberw} for an overview.) The first of these, which has
inspired much of the later work, was a study of the ground state
energy of a Bose gas with repulsive interaction at
low density $\rho$. Per unit volume, it is given by
\begin{equation}\label{e0a}
e_0(\rho) = 4\pi a \rho^2 + o(\rho^2) \qquad {\rm for\ } a^3\rho\ll 1 
\end{equation}
in three spatial dimensions.  Here, $a>0$ denotes the {\it scattering
  length} of the interparticle interaction, and units are chosen such
that $\hbar=2m=1$, with $m$ the mass of the particles. A lower bound
on $e_0(\rho)$ of the correct form (\ref{e0a}) was proved by Lieb and
Yngvason in \cite{LY1998}. Much earlier, Dyson \cite{dyson} had
already proved an upper bound of the desired form, at least in the
special case of hard-sphere particles. An extension of his calculation
to arbitrary repulsive interaction potentials was given in
\cite{LSY00}.

The methods introduced in \cite{LY1998} have been extended to treat
the case of fermions as well, for the study of both the ground state
energy \cite{LSSfermi} and the free energy at positive temperature
\cite{FermiT}. We are concerned here with the extension of (\ref{e0a})
to positive temperature, at least as far as a lower bound is
concerned.  That is, our goal is to derive a lower bound on the {\it
   free energy} of a dilute Bose gas at density $\rho$ and temperature
$T$. Much of the complication in such an estimate is caused by the
existence of a {\it Bose-Einstein condensate} for temperatures below
some critical temperature.  Although the existence of a condensate for
interacting Bose gases has so far eluded a mathematical proof, its
presence can easily be shown in the case of non-interacting particles.
A short review of the Bose gas without interaction among the particles
is given in Subsection~\ref{ideals} below.

One of the main ingredients in our estimate is a method to quantify
correlations present in the state of the interacting system. This
method has been introduced in \cite{RSjellium}; it does not
immediately apply below the critical temperature for Bose-Einstein
condensation, however. We have been able to overcome this difficulty
with the aid of coherent states.

\subsection{Definition of the Model}

We consider a system of $N$ bosons, confined to a three-dimensional
flat torus of side lengths $L$, which we denote by $\Lambda$. The
one-particle state space is thus $L^2(\Lambda,dx)$, and the Hilbert
space for the system is the {\it symmetric} $N$-fold tensor product
$\Hh_N= L^2_{\rm sym}(\Lambda^N,d^Nx)$, i.e., the space of square
integrable functions of $N$ variables that are invariant under
exchange of any pair of variables. The {\it Hamiltonian} is given as
\begin{equation}\label{hamori}
H_N = \sum_{i=1}^N -\Delta_i + \sum_{1\leq i<j\leq N} v(d(x_i,x_j))\,.
\end{equation}
Here, $\Delta$ denotes the Laplacian on $\Lambda$, and $d(x,y)$
denotes the distance between points $x$ and $y$ on the torus
$\Lambda$.  The particle interaction potential $v:\R_+\mapsto \R_+
\cup \{\infty\}$ is assumed to be a non-negative and measurable
function. It is allowed to take the value $+\infty$ on a set of
positive measure, corresponding to hard sphere particles. In this
case, the domain of the Hamiltonian has to be suitably restricted to
functions that vanish on the set where the interaction potential is
infinite. We assume that $v$ has a finite range $R_0$, i.e., $v(r) =
0$ for $r> R_0$. In particular, it has a finite {\it scattering
   length}, which we denote by $a$. We will recall the definition
of $a$ in Subsection~\ref{defas} below.

We note that in a concrete realization of $\Lambda$ as the set
$[0,L]^3\subset \R^3$, $\Delta$ is the Laplacian on $[0,L]^3$ with
{\it periodic} boundary conditions. Moreover, the distance $d(x,y)$ is
given as $d(x,y)=\min_{k\in \Z^3} |x-y-k L|$. Note also that
$v(d(x,y))= \sum_{k\in\Z^3} v(|x-y-kL|)$ if $L> 2R_0$.

The {\it free energy} (per unit volume) of the system at inverse
temperature $\beta= 1/T >0$ and density $\rho>0$ is given by
\begin{equation}\label{deff}
f(\beta,\rho) = -\frac 1\beta \lim \frac {1}{|\Lambda|} 
\ln \Tr_{\Hh_N} \exp\left(
   -\beta H_N\right)\,,
\end{equation}
where $\lim$ stands for the usual thermodynamic limit $L\to\infty$,
$N\to\infty$ with $\rho=N/|\Lambda|$ fixed. Here, we denote the volume of
$\Lambda$ by $|\Lambda|=L^3$. Existence of the thermodynamic limit in
(\ref{deff}) can be shown by standard methods, see, e.g., \cite{rob,rue}.

We are interested in a bound on $f$ in the case of a {\it dilute} gas,
meaning that $a^3\rho$ is small. The dimensionless parameter
$\beta\rho^{2/3}$ is of order one (or larger), however.  Note that
sometimes in the literature the case of small $\rho$, but fixed $a$
and $\beta$, is understood with the term \lq\lq dilute\rq\rq. This
corresponds to a high-temperature (classical) limit and is not what we
want to study here.

\subsection{Ideal Bose Gas}\label{ideals}

In the case of vanishing interaction potential ($v\equiv 0$), the free
energy can be evaluated explicitly. It is given as
\begin{equation}\label{fnodeb}
f_0(\beta,\rho) = \sup_{\mu\leq 0} 
\left\{ \mu \rho+ \frac 1{(2\pi)^3\beta} \int_{\R^3}
dp\, \ln\left( 1 - e^{-\beta (p^2- \mu)}\right) \right\}\,.
\end{equation}
The supremum is uniquely attained at $\mu=\mu_0(\beta,\rho)= d/d\rho\,
f(\beta,\rho)\leq 0$. 
If $\rho$ is bigger than the {\it critical density} 
\begin{equation}
\rho_c(\beta)\equiv \frac 1{(2\pi)^3} \int_{\R^3} dp\, \frac 1 {e^{\beta
     p^2} -1} = 
(4\pi\beta)^{-3/2}\sum_{\ell\geq
   1}\ell^{-3/2}\,,
\end{equation}
the supremum is attained at $\mu_0=0$, whereas for
$\rho < \rho_c(\beta)$, it is attained at some
$\mu_0= \mu_0(\beta,\rho)<0$. In particular, 
\begin{equation}\label{eq16}
\frac 1{(2\pi)^3} \int_{\R^3} dp\, \frac 1 {e^{\beta
     (p^2 - \mu_0)} -1} = \min\{ \rho, \rho_c(\beta)\}\,.
\end{equation}

Note also that the scaling relation $f_0(\beta,\rho)= \rho^{5/3}
f_0(\beta\rho^{2/3},1)$ holds for an ideal Bose gas. In particular,
the dimensionless quantity $\beta\rho^{2/3}$ is the only relevant
parameter.

\subsection{Scattering Length}\label{defas}

The scattering length of a potential $v$ can be defined as follows
(see Appendix~A in \cite{LY2d}, or Appendix~C in \cite{oberw}): For
$R\geq R_0$,
\begin{align}\nonumber
\frac{ 4\pi a}{1-a/R} = &\inf \left\{  \int_{|x|\leq R} dx\, \left(
     |\nabla\phi(|x|)|^2 + \half v(|x|) \phi(|x|)^2\right) \, : \right. \\ &
   \left. \phantom{\int_{\R^3}} \qquad \phi:[0,R]\mapsto \R_+\,,\,
   \phi(R) = 1\right\}\,.\label{defsl}
\end{align}
For this definition to make sense, $v$ need not necessarily be
positive, one only has to assume that $-\Delta + \half v$ (as an
operator on $L^2(\R^3)$) does not have any negative spectrum. We will
restrict our attention to non-negative $v$, however. The infimum in
(\ref{defsl}) is attained uniquely. Moreover, the minimizer has a
trivial dependence on $R$: for some function $\phi_{v}$ (independent
of $R$) it can be written as $\phi_{v}(|x|)/\phi_v(R)$. Note that $a$
is independent of $R$, and also that $\phi_v(|x|)= 1-a/|x|$ for
$|x|\geq R_0$.

\subsection{Main Theorem}

Our main result is the following lower bound on the free energy, 
defined in (\ref{deff}). It gives a bound on the leading order
correction, compared with a non-interacting gas, in the case of small
$a^3\rho$ and fixed $\beta\rho^{2/3}$.

\begin{thm}[Lower bound on free energy of dilute Bose gas]\label{T1}
There is a function $C:\R_+\mapsto \R_+$, uniformly bounded on bounded
subsets of $\R_+$, and an $\alpha > 0$ such  that 
\begin{equation}\label{fmt}
f(\beta,\rho) \geq f_0(\beta,\rho) + 4\pi a \left( 2 \rho^2 -
   [\rho-\rho_c(\beta)]_+^2\right) \big(1 - o(1)\big) \,,
\end{equation}
with 
\begin{equation}
o(1) \leq C\big( (\beta\rho^{2/3})^{-1} \big) (a\rho^{1/3})^\alpha\,.
\end{equation}
\end{thm}

Here, $[\,\cdot\,]_+=\max\{\,\cdot\,,0\}$ denotes the positive
part. In the case of non-interacting particles, the expression
$[\rho-\rho_c(\beta)]_+$ is just the {\it condensate density}.
\bigskip

\noindent {\bf Remarks.}
\begin{itemize}
  
\item [1.] Since $C(t)$ is uniformly bounded for bounded $t$, our
  estimate is uniform in the parameter $(\beta\rho^{2/3})^{-1}$ as
  long as it stays bounded. I.e., our result is uniform in the
  temperature for temperatures not much greater than the critical
  temperature (for the non-interacting gas). In particular, we recover
  the result (\ref{e0a}) in the zero temperature limit. The error term
  is worse, however; in \cite{LY1998}, it was shown that the exponent
  $\alpha$ can be taken to be $\alpha = 3/17$ at $T=0$, whereas our
  proof shows that $\alpha$ can be chosen slightly larger than
  $0.00087$ (independent of $T$). This value has no physical
  significance, however, it merely reflects the multitude of estimates
  needed to arrive at our result.

\item [2.]  The error term, $o(1)$, in our lower bound depends on the
   interaction potential $v$, besides its scattering length $a$, only
   through its range $R_0$. This dependence could in principle be
   displayed explicitly. By cutting off the potential in a suitable
   way, one can then extend the result to infinite range potentials
   (with finite scattering length). See Appendix~B in \cite{LSY00} for
   details.
   
 \item [3.] For $\rho\leq \rho_c(\beta)$ (i.e., above the critical
   temperature), the leading order correction term is given by $8\pi a
   \rho^2$, compared with $4\pi a\rho^2$ at zero temperature. The
   additional factor 2 is an exchange effect; heuristically speaking,
   it is a result of the symmetrization of the wave functions. This
   symmetrization only plays a role if the particles are in different
   one-particle states, which they are essentially always above the
   critical temperature. Below the critical temperature, however, a
   macroscopic number of particles occupies the zero-momentum state;
   there is no exchange effect among these particles, which explains the
   subtraction of the square of the condensate density in (\ref{fmt}).

\item [4.]  We note that $f_0(\beta,\rho)$ has a discontinuous
   third derivative with respect to $\rho$ at $\rho=\rho_c(\beta)$ or,
   equivalently, a discontinuous third derivative with respect to
   $T=1/\beta$ at the critical temperature. Since the specific heat
   $c_V(\beta,\rho)$ can be expressed in terms of the free energy as
   $c_V(\beta,\rho)= - T d^2/(dT)^2 \, f(\beta,\rho)$, it has a
   discontinuous derivative (with respect to $T$) at the critical
   temperature. The first order correction term in (\ref{fmt}) has a
   discontinuous second derivative at this value. Considering only this
   term and neglecting higher order corrections, this would mean that
   the specific heat is actually discontinuous at the critical
   temperature.

\item [5.]  Our method applies also to particles with internal degrees
   of freedom, e.g., to particles with nonzero spin. For simplicity, we
   treat only the case of spinless particles here.

\item [6.] Although we provide only a lower bound in this paper, one
   can expect that the second term in (\ref{fmt}) gives the correct
   leading order correction to the free energy (see, e.g.,
   \cite[Chapter~12.4]{huang}). To prove this, one has to derive an
   appropriate upper bound on $f(\beta,\rho)$, which has not yet been
   achieved, however.  We note that a naive upper bound using first
   order perturbation theory yields (\ref{fmt}) with $4\pi a$ replaced
   by $\half \int dx\, v(|x|)$, which need not be finite, however (and
   is always strictly greater than $4\pi a$).

\end{itemize}

The remainder of this paper is devoted to the proof of
Theorem~\ref{T1}. The proof is quite lengthy and is split into several
subsections. To guide the reader, we start every subsection with a
short summary of what will be accomplished.

\bigskip
\noindent {\it Acknowledgments.} It is a pleasure to thank Elliott
Lieb and Jan Philip Solovej for many inspiring discussions.

\numberwithin{equation}{subsection}

\section{Proof of Theorem~\ref{T1}}

In the following, we find it convenient to think of $\Lambda$ as the
set $[0,L]^3$ embedded in $\R^3$. We will also assume $L$ to be
large. In particular, $L> 2R_0$, but $L$ will also be assumed to be large
compared with several other parameters (with are independent of $L$)
appearing below. This is justified since we are only interested in
quantities in the thermodynamic limit $L\to \infty$.

In many places in our proof, the Heaviside step function $\theta$ will
appear. We point out that we use the convention that $\theta$ equals
$1$ at the origin, i.e., $\theta(t)=0$ for $t<0$, and $\theta(t)=1$
for $t\geq 0$.

\subsection{Reduction to Integrable Potentials}

Recall that we do not want to restrict our attention to interaction
potentials that are integrable. For the Fock space treatment in the
next subsection, it will be necessary that $v$ has finite Fourier
coefficients, however. As a first step, we will therefore replace the
interaction potential $v$ by a smaller potential $\widetilde v$ whose
integral is bounded by some number $8\pi \varphi$. The scattering
length of the new potential will be smaller than $a$, however. In the
following lemma, we show that as long as $\varphi$ is much greater
than $a$, the change in the scattering length remains small.

\begin{lem} Let $v:\R_+ \to \R_+ \cup \{\infty\}$ have finite
  scattering length $a$. For any $\eps >0$, there exists a $\widetilde
  v$, with $0\leq \widetilde v(r) \leq v(r)$ for all $r$, such that
  $\int_0^\infty dr\, r^2 \widetilde v(r) \leq 2 \varphi$, and such
  that the scattering length of $\widetilde v$, denoted by $\widetilde
  a$, satisfies
\begin{equation}\label{eqlem1}
\widetilde a \geq a  \left( 1 - \sqrt{a/\varphi}\right)\left(1-\eps\right)\,.
\end{equation}
\end{lem}

\begin{proof}
   Without loss of generality, we may assume that $\varphi > a$.  Let
   $R = \inf\{ s\, : \, \int_s^\infty dr\, r^2 v(r) < \infty\}$. We
   note that $R$ is finite; in fact $R\leq a$. This follows from the
   fact that $2 a \geq \int_0^\infty dr\, r^2 v(r)|\phi_v(r)|^2$, where
   $\phi_v$ denotes the minimizer of (\ref{defsl}) (for $R=\infty$), as
   introduced in Subsection~\ref{defas}. Since it satisfies
   $\phi_v(r)\geq 1-a/r$ (see Appendix~B in \cite{LY2d}),
   $\int_s^\infty dr\, r^2 v(r)$ is finite for $s>a$.

Assume first that $\int_R^\infty dr\, r^2 v(r) \geq 2\varphi$. 
The function $s\mapsto \int_s^\infty dr\, r^2 v(r)$ is continuous for
$s>R$. We can thus choose $s\geq R$ such that $\int_s^\infty dr\, r^2
v(r) = 2 \varphi$, and $\widetilde v(r) = v(r) \theta (r-s)$.

To obtain an upper bound on $a$, we can use a trial function $\phi(r)
= (\phi_{\widetilde v}(r) - \phi_{\widetilde v}(s) s/r)\theta(r-s)$ in
the variational principle (\ref{defsl}). We note that $\phi$ is a
non-negative function, since $\phi_{\widetilde v}(r)$ is monotone
increasing in $r$ \cite{LY2d}. By partial integration, using
the variational equation $-\Delta \phi_{\widetilde v}(|x|)+\half
\widetilde v(|x|)\phi_{\widetilde v}(|x|)=0$, we have 
\begin{align}\label{argule}
4\pi a  & \leq \int_{\R^3} dx \left( |\nabla \phi(|x|)|^2 + \half v(|x|)
   |\phi(|x|)|^2 \right) \\ &  = 4\pi \left( \widetilde a + s
   \phi_{\widetilde v}(s)\right) + 2\pi \phi_{\widetilde v}(s)
\int_s^\infty dr\, r^2 v(r) \frac sr \left( \phi_{\widetilde v}(s)
   \frac sr  - \phi_{\widetilde v}(r) \right)\,. \nonumber
\end{align}
The last
term is negative and can be dropped for an upper bound. To obtain an
upper bound on $s\phi_{\widetilde v}(s)$, we note that
$\phi_{\widetilde v}(s) \geq 1 - \widetilde a/s$, and hence
\begin{equation}
s \phi_{\widetilde v}(s)  \leq \frac{\widetilde a }{
   1/\phi_{\widetilde v}(s) - 1}\,.
\end{equation}
For an upper bound on $\phi_{\widetilde v}(s)$, 
we use again the monotonicity of  $\phi_{\widetilde v}(r)$, which
allows us to estimate 
\begin{equation}\label{samar}
a \geq \widetilde a \geq \half \int_s^\infty dr\, r^2 v(r)
\phi_{\widetilde v}(r)^2 \geq  \phi_{\widetilde v}(s)^2 \varphi\,.
\end{equation}
This yields $\phi_{\widetilde v}(s) \leq \sqrt{a/\varphi}$.

Altogether, we have thus shown that
\begin{equation}
a \leq \widetilde a + s \phi_{\widetilde v}(s) \leq 
\widetilde a \left( 1 + \frac
   1{ \sqrt{\varphi/a} -1}\right)\,.
\end{equation}
This proves (\ref{eqlem1}) (with $\eps = 0$) under the assumption that
$\int_R^\infty dr\, r^2 v(r) \geq 2\varphi$.

Consider now the case when $\int_R^\infty dr\, r^2 v(r) = 2\varphi -
T$ for some $T>0$. If $R=0$, we can take $\widetilde v=v$, and there
is nothing to prove. Hence we can assume that $R>0$. By definition, we
have that $\int_{R(1-\eps)}^R dr\, r^2 v(r) = \infty$ for any
$\eps>0$. Hence there exists a $\tau$ (depending on $T$ and $\eps$)
such that $\int_{R(1-\eps)}^R dr\, r^2 \min\{v(r),\tau\} = T$. We can
then take
\begin{equation}
\widetilde v(r) = \left\{ \begin{array}{ll}
v(r) & {\rm for\ } r \geq R\\
\min\{v(r),\tau\} & {\rm for\ } (1-\eps)R \leq r < R\\
0 & {\rm otherwise \,.}
\end{array}\right.
\end{equation} 
Applying the same argument as in (\ref{argule}), with $s= R$, we have
$a\leq \widetilde a + R\phi_{\widetilde v}(R)$. Now $R\leq a$, and
$\phi_{\widetilde v}(R(1-\eps)) \leq \sqrt{a/\varphi}$ using the same
argument as in (\ref{samar}), noting that $\widetilde v(r)=0$ for
$r\leq R(1-\eps)$. Moreover, since $|\nabla \phi_{\widetilde
  v}(|x|)|\leq \widetilde a /|x|^2$, as shown in
\cite[Eq.~(3.33)]{LSY00}, $|\phi_{\widetilde
  v}(R(1-\eps))-\phi_{\widetilde v}(R)| \leq \eps \widetilde a
R^{-1}/(1-\eps)$, and thus
\begin{equation}
a \leq \widetilde a \frac{1}{1-\eps} + a \sqrt{\frac a\varphi}\,.
\end{equation}
This finishes the proof of the lemma.
\end{proof}

As an example, consider the case of a pure hard sphere interaction,
i.e., $v(r) = \infty$ for $r\leq a$, and $v(r)=0$ for $r>a$. In this
case, we can choose $\widetilde v(r)= 6\varphi a^{-3} \theta(a-r)$.
The scattering length of $\widetilde v$ is given by $\widetilde a = a
( 1- \sqrt{a/(6\varphi)} \tanh \sqrt{6\varphi/a})$ in this case. Note
that $\tanh t \leq 1$ for all $t$. In particular, $\widetilde a \geq
a(1-\sqrt{a/(6\varphi)})$.

For a lower bound, we can simply replace $v$ by $\widetilde v$,
i.e, we have $H_N\geq \widetilde H_N$, with
\begin{equation}\label{ham}
\widetilde H_N = \sum_{i=1}^N -\Delta_i + \sum_{1\leq i<j\leq N}
\widetilde v(d(x_i,x_j))\,.
\end{equation}
If we choose $\eps \leq \sqrt{a/\varphi}$, the error in the scattering
length $\widetilde a$ is of the order $\sqrt{a/\varphi}$. We will
choose $\varphi \gg a$ below.

\subsection{Fock Space}

In the following, it will be convenient to give up the restriction on
the particle number and work in Fock space instead. This has the
advantage that the problem of condensation can be dealt with with the
aid of coherent states, which will be introduced in the next
subsection. Our treatment differs slightly from the usual
grand canonical ensemble since we do not simply introduce a chemical
potential as Lagrange multiplier to control the number of particles,
but we add a quadratic expression in $N$ to the Hamiltonian. This
gives a bitter control on the particle number.

Let $\mu_0\leq 0$ be the chemical potential of the ideal Bose gas,
which is the quantity that maximizes the expression in (\ref{fnodeb}).
Let $\Ff= \bigoplus_N \Hh_N$ be the bosonic Fock space over
$L^2(\Lambda)$.  Let $\ad{p}$ and $\an{p}$ denote the usual creation
and annihilation operators of plane waves in $\Lambda$ with wave
functions $L^{-3/2}e^{-ipx}$.  We define a Hamiltonian $\dH$ on Fock
space as
\begin{equation}\label{defhf}
\dH=\dT+\dV+\dK + \mu_0 N \,,
\end{equation}
with
\begin{equation}
\dT= \sum_{p} \big(p^2 -\mu_0\big) \ad{p}\an{p}\quad , \qquad \dV= \frac
1{2|\Lambda|} \sum_{p,k,l} \widehat v(p)
\ad{k+p}\ad{l-p}\an{k}\an{l}\,,
\end{equation}
and
\begin{equation}\label{defdK}
\dK= 4\pi \widetilde a\frac {C}{|\Lambda|} \left( \dN - N\right)^2\,.
\end{equation}
Here and in the following, all sums are over $p\in \frac{2\pi}L \Z^3$.
The Fourier transform of $\widetilde v$ is denoted by $\widehat v$,
i.e., $\widehat v(p) = \int_\Lambda dx\, \widetilde v(|x|)
e^{-ipx}$. It is uniformly bounded; in fact, $|\widehat v(p)|\leq \widehat
v(0) \leq 8\pi \varphi$, where $\varphi$ was introduced in the
previous subsection.  The number operator $ \sum_{p} \ad{p}\an{p} $ is
denoted by $\dN$, whereas $N$ is just a parameter.  The parameter $C$
is positive and will be chosen later on.

The Hamiltonian $\dH$ commutes with the number operator $\dN$, and can
be thought of a direct sum of its restrictions to definite particle
number. Note that the restriction to particle number $N$ is just
$\widetilde H_N$, i.e, $\dH = \widetilde H_N$ on the sector of
particle number $N$. This implies, in particular, that
\begin{equation}
\Tr_{\Hh_N} \exp\big( -\beta \widetilde H_N\big) \leq \Tr_{\Ff} \exp\left(
   -\beta \dH\right)\,.
\end{equation}
We will proceed deriving an upper bound on the latter expression.

\subsection{Coherent States}

To obtain an upper bound on the partition function $\Tr_\Ff\,
\exp(-\beta \dH)$, we use the method of coherent states \cite{lsybog}.
Effectively, this replaces the operators $\ad{p}$ and $\an{p}$ by
numbers. This can be viewed as a rigorous version of part of the
Bogoliubov approximation, where one replaces the operators $\ad{0}$
and $\an{0}$ by numbers. Such a replacement is particularly useful if
the zero-mode is \lq\lq macroscopically occupied\rq\rq, i.e, if
$\ad{0}\an{0} \sim |\Lambda|$. We will use this method not only for
$p=0$, however, but for a whole range of momenta $|p|< p_c$ for some
$p_c\geq 0$. Although not macroscopic, their occupation will be large
enough to require this separate treatment.

To be more precise, let us pick some $p_c\geq 0$ and write
$\Ff=\Ff_<\otimes \Ff_>$, where $\Ff_<$ and $\Ff_>$ denote the Fock
spaces corresponding to the modes $|p|< p_c$ and $|p|\geq p_c$,
respectively.  Let $M$ denote the number of $p\in \frac{2\pi}L \Z^3$
with $|p|< p_c$.  As shown in \cite{lsybog}, the Berezin-Lieb
inequality \cite{ber,lie} implies that
\begin{equation}\label{bli}
\Tr_\Ff\, \exp(-\beta \dH) \leq \int_{\C^M} d^{M}\! z\, \Tr_{\Ff_>}\,
\exp\big(-\beta \dH^{\rm s}(\vec z)\big) \,.
\end{equation}
Here, $\vec z$ denotes the vector $(z_1,\dots,z_M)\in \C^M$, $d^M\!z =
\prod_{i=1}^M dz_i$ and $dz= \pi^{-1} dx\, dy$ with $x=\Re(z)$,
$y=\Im(z)$.
Moreover,
$\dH^{\rm s}(\vec z)$ is the {\it upper symbol} of
the operator $\dH$. It is an operator on $\Ff_>$, parametrized by
$\vec z$, and can be written in the
following way. Let $|\vec z\rangle \in \Ff_<$ denote the {\it coherent
state} 
\begin{equation}
|\vec z\rangle = \exp\left(\mbox{$\sum_{|p|< p_c}$} z_p \ad{p} -
   z_p^*
   \an{p}\right)|0\rangle \equiv U(\vec z) |0\rangle\,,
\end{equation}
with $|0\rangle$ the vacuum in the Fock space $\Ff_<$. Then the {\it
  lower symbol} of $\dH$ is given by $\dH_{\rm s}(\vec z)= \langle
\vec z|\dH|\vec z\rangle$. Since $\an{p}|\vec z\rangle = z_p |\vec
z\rangle$, the lower symbol is obtained from the expression
(\ref{defhf}) by simply replacing all the $\an{p}$ by $z_p$ and the
$\ad{p}$ by $z_p^*$ for all $|p|<p_c$. The upper symbol can be
obtained from the lower symbol by replacing $|z_p|^2$ by $|z_p|^2-1$,
for instance, and similarly with other polynomials in $z_p$; see,
e.g., \cite{lsybog} for details. We can then write $\dH^{\rm s}(\vec
z)$ in the following way.  Denoting by $\dN_{\rm s}(\vec z)= |\vec
z|^2 + \sum_{|p|\geq p_c} \ad{p}\an{p}$ the lower symbol of the number
operator, we have
\begin{equation}\label{uplo}
\dH^{\rm s}(\vec z) = \dH_{\rm s}(\vec z) - \Delta \dH(\vec z)\,,
\end{equation}
with 
\begin{align}\nonumber
  \Delta \dH(\vec z) = & \sum_{|p|< p_c} \big(p^2 -\mu_0\big) + \frac
  1{2|\Lambda|}\Biggl[ \widehat v(0)\left( 2 M \dN_s(\vec z) -
    M^2\right) \\ \nonumber &+ 2 \sum_{|l|< p_c,\, |k|\geq p_c}
  \widehat v(l-k) \ad{k}\an{k} + \sum_{|l|< p_c,\, |k|< p_c} \widehat
  v(l-k) \left(2 |z_k|^2 -1\right)\Biggl] \\ & +\frac{4\pi \widetilde
    a C}{|\Lambda|} \left[ 2|\vec z|^2 + M ( 2 \dN_{\rm s}(\vec z) - 2
    N - M ) \right]\,.\label{twl}
\end{align}
Here, we have used that $\widehat v(p)= \widehat v(-p)$.

Since $\widetilde v$ is a non-negative function, $|\widehat v(p)|\leq \widehat
v(0) \leq 8\pi \varphi$ for all $p$. Hence we obtain the bound 
\begin{equation}\label{dhb}
\Delta \dH(\vec z) \leq M \big( p_c^2 -\mu_0\big) +
\frac{16\pi\varphi}{|\Lambda|} M \dN_s(\vec z) + \frac{8\pi \widetilde
   a C}{|\Lambda|} \left[
   |\vec z|^2 +  M ( \dN_{\rm s}(\vec z) - N ) \right]\,.
\end{equation}
(Here, we have used again the positivity of $\widetilde v$.) 
Denoting by $\dK_{\rm s}(\vec z)=\langle \vec z|\dK|\vec
z\rangle$ the lower symbol of $\dK$ (and, similarly, for $\dT$ and
$\dV$ below), we have 
\begin{equation}\label{ksr}
\dK_{\rm s}(\vec z) = \frac{4\pi \widetilde a C}{|\Lambda|}
\left( \left( \dN_{\rm
       s}(\vec z) -N \right)^2 + |\vec z|^2\right) \geq  \frac{4\pi
   \widetilde a C}{|\Lambda|} \left( \dN_{\rm
       s}(\vec z) -N \right)^2\,.
\end{equation}
We can use part of $\dK_{\rm s}(\vec z)$ to estimate $-\Delta \dH$ from
below independently of $\vec z$. More precisely, we have
\begin{align}\nonumber
& \half \dK_{\rm s}(\vec z) - \Delta \dH(\vec z)  \\ \nonumber & \geq
-M \big(p_c^2 - \mu_0\big) - \frac
{8\pi N}{|\Lambda|} (2\varphi M + \widetilde a C) - 32\pi \widetilde a C
\frac{(M+1)^2}{|\Lambda|} \left(1 + \frac {2\varphi}{\widetilde a C}\right)^2
\\ \label{defz1}
& \equiv -Z^{(1)} \,.
\end{align}
Note that $M\sim p_c^3 |\Lambda|$ in the thermodynamic limit. We will
choose the parameters $p_c$, $\varphi$ and $C$ such that $Z^{(1)} \ll
|\Lambda|  a \rho^2$ for small $a\rho^{1/3}$.

With the definition
\begin{equation}
  F_{\vec z}(\beta) \equiv -\frac 1\beta \ln\Tr_{\Ff_>}\,
\exp\big(-\beta(\dT_{\rm s}(\vec z)+ \dV_{\rm s}(\vec z)+ \half \dK_{\rm
   s}(\vec z))\big)\,,
\end{equation}
(\ref{bli}) and the estimates above imply that 
\begin{equation}\label{eq234}
-\frac 1\beta \ln \Tr_\Ff\, \exp(-\beta \dH) \geq  \mu_0 N 
-\frac 1\beta \ln \int_{\C^M}
d^{M}\! z\, \exp\big(-\beta F_{\vec z}(\beta)\big) - Z^{(1)}
\,.
\end{equation}
Hence it remains to derive a lower bound on $F_{\vec z}(\beta)$.

Let $\Gamma_{\vec z}$ denote the Gibbs state of $\dT_{\rm s}(\vec z)+
\dV_{\rm s}(\vec z)+ \half \dK_{\rm s}(\vec z)$ on $\Ff_>$, for
inverse temperature $\beta$. Let $\Pi_0=|0\rangle\langle 0|$ denote
the vacuum state in $\Ff_<$. Denoting by $\Upsilon^{\vec z}$ the state
$\Upsilon^{\vec z}\equiv U(\vec z)\Pi_0 U(\vec z)^\dagger \otimes
\Gamma_{\vec z}$ on the full Fock space $\Ff$, we can write
\begin{equation}\label{fzb}
F_{\vec z}(\beta) = \Tr_\Ff\big[ \big(\dT+\dV+\half \dK\big)
\Upsilon^{\vec z}\big] - \frac 1\beta S(\Upsilon^{\vec z})\,.
\end{equation}
Here, $S(\Gamma)= -\Tr_\Ff\, \Gamma \ln \Gamma$ denotes the
{\it von-Neumann entropy}.

\subsection{Relative Entropy and A Priori Bounds}\label{secas}

In the following, we want to derive a lower bound on $F_{\vec
  z}(\beta)$. Although we do not have an upper bound available, we can
assume an appropriate upper bound without loss of generality; if the
assumption is not satisfied, there is nothing to prove (as far as a
lower bound in concerned). This upper bound can be formulated as a
bound on the {\it relative entropy} between the state $\Upsilon^{\vec
  z}=U(\vec z)\Pi_0 U(\vec z)^\dagger \otimes \Gamma_{\vec z}$ defined
above and a simple reference state (describing non-interacting
particles).  Together with a bound on the total number of particles,
this estimate on the relative entropy contains all the information we
need in order to prove the desired properties of the state
$\Upsilon^{\vec z}$ that will allow us to derive a lower bound on
(\ref{fzb}).

We note that, for any state $\Gamma$ of the form $\Gamma=U(\vec
z)\Pi_0 U(\vec z)^\dagger\otimes \Gamma^>$ for some state $\Gamma^>$
on $\Ff_>$, 
\begin{align}\nonumber
  \Tr_\Ff\big[ \dT \Gamma\big] - \frac 1\beta S(\Gamma)  &=
  \Tr_{\Ff_>}\big[ \dT_{\rm s}(\vec z) \Gamma^> \big] - \frac 1\beta
  S(\Gamma^>) \\ & \geq -\frac 1\beta  \ln \Tr_{\Ff_>}\,
\exp\big(-\beta \dT_{\rm s}(\vec z)\big)\,.\label{nio}
\end{align}
In fact, the difference between the right and left sides of
(\ref{nio}) is given by $\beta^{-1}S(\Gamma,\Omega_0^{\vec z})$, where
$S$ denotes the {\it relative entropy}.  For two general states
$\Gamma$ and $\Gamma'$ on Fock space, it is given by
\begin{equation}\label{defS}
S(\Gamma, \Gamma') = \Tr_\Ff\, \Gamma\big( \ln \Gamma - \ln
\Gamma'\big)\,.
\end{equation}
Note that the relative entropy is a non-negative quantity. The state
$\Omega_0^{\vec z}$ is given by $\Omega_0^{\vec z} = U(\vec z)\Pi_0
U(\vec z)^\dagger \otimes \Gamma^0$, where $\Gamma^0$ is the Gibbs
state of $\dT_{\rm s}(\vec z)$ on $\Ff_>$ (which is independent of
$\vec z$).

For $\Gamma=\Upsilon^{\vec z}$, we have
\begin{equation}
S(\Upsilon^{\vec z},\Omega_0^{\vec z}) 
= \Tr_{\Ff_>} \, \Gamma_{\vec z}\big( \ln
\Gamma_{\vec z} - \ln
\Gamma^0\big) = S(\Gamma_{\vec z},\Gamma^0)\,. \label{defg0}
\end{equation}
From these considerations, together with the positivity of $\dV$, we
conclude that (\ref{fzb}) is bounded from below by 
\begin{equation}
  F_{\vec z}(\beta)  \geq - \frac 1 \beta \ln \Tr_{\Ff_>}\,
\exp\big(-\beta \dT_{\rm s}(\vec z)\big) + 
\half \Tr_\Ff\big[\dK \Upsilon^{\vec z}\big] 
+ \frac 1 \beta S(\Gamma_{\vec z},\Gamma^{0})\,.
\end{equation}
Hence we can distinguish the following two cases:

\begin{itemize}
\item[A)] The following lower bound on $F_{\vec z}(\beta)$ holds:
\begin{equation}\label{lowf}
F_{\vec z}(\beta) \geq - \frac 1\beta \ln \Tr_{\Ff_>}\,
\exp\big(-\beta \dT_{\rm s}(\vec z)\big)+ 8\pi
|\Lambda| \widetilde a\rho^2\,.
\end{equation}
\item[B)] Inequality (\ref{lowf}) is false, in which case
\begin{equation}\label{asss}
S(\Gamma_{\vec z},\Gamma^{0})\leq 8\pi |\Lambda| \widetilde a\beta\rho^2
\end{equation}
and
\begin{equation}\label{assn}
\Tr_\Ff\big[\dK \Upsilon^{\vec z}\big] \leq 16 \pi |\Lambda|
\widetilde a
\rho^2\,.
\end{equation}
\end{itemize}

From now on, will consider case B, i.e., we will assume (\ref{asss})
and (\ref{assn}) to hold. The lower bound we will derive on $F_{\vec
  z}(\beta)$ below will actually be worse then the bound (\ref{lowf})
above; i.e., the bound in case B holds in any case, irrespective of
whether the assumptions (\ref{asss}) and (\ref{assn}) actually hold.

Although the relative entropy does not define a metric, it measures
the difference between two states in a certain sense. In particular,
it dominates the trace norm \cite[Thm.~1.15]{ohya}:
\begin{equation}\label{sn}
S(\Gamma,\Gamma') \geq \half \| \Gamma-\Gamma'\|_1 \,.
\end{equation}
This inequality is a special case of the fact that the relative
entropy decreases under completely positive trace-preserving (CPT)
maps. In fact, inequality (\ref{sn}) can be obtained using monotonicity
under the CPT map $\Gamma \mapsto \Tr_\Ff[P \Gamma] \oplus
\Tr_\Ff[(1-P) \Gamma]$, where $P$ is the projection onto the subspace
where $\Gamma-\Gamma' \geq 0$.

Although we have the upper bound (\ref{asss}) on the relative entropy,
inequality (\ref{sn}) is of no use for us since the relative entropy
is of the order of the volume of the system, while the right side of
(\ref{sn}) never exceeds 2. To make use of (\ref{sn}), we must not
look at the state on the full Fock space (over the whole volume) but
rather on its restriction to a small subvolume. We do this in
Subsection~\ref{sect:locrel} below. Again, the monotonicity of the
relative entropy will be used in an essential way.

We note that (\ref{assn}) implies the following simple upper bound on
$|\vec z|^2$. From (\ref{defdK}) and (\ref{assn}), 
\begin{equation}\label{boun}
|\vec z|^2 - N \leq \Tr_\Ff\big[ ( \dN - N)\Upsilon^{\vec z}\big ]
\leq \left(
\Tr_\Ff\big[(\dN-N)^2 \Upsilon^{\vec z}\big]\right)^{1/2} \leq
\frac 2 {\sqrt{C}} |\Lambda| \rho\,,
\end{equation}
and hence
\begin{equation}\label{zap}
\rho_{\vec z} \equiv \frac{|\vec z|^2}{|\Lambda|} \leq \rho \left( 1 +
   \frac 2{\sqrt{C}}\right)\,.
\end{equation}
We will choose $C\gg 1$ below.

\subsection{Replacing Vacuum}\label{replsec}

In the following, we want to derive a lower bound on the expectation
value of the interaction energy $\dV$ in the state $\Upsilon^{\vec
   z}$, i.e., on 
\begin{equation}\label{s4e}
\Tr_\Ff\big[ \dV \Upsilon^{\vec z}\big]  = \Tr_{\Ff_>}\big[  \dV_{\rm s}(\vec
   z)\Gamma_{\vec z}\big]  \,.
\end{equation}
For reasons that will be explained later (see
Subsection~\ref{relesec}), we find it necessary to replace the vacuum
$\Pi_0$ on $\Ff_<$ in the definition of the state $\Upsilon^{\vec
  z}=U(\vec z)\Pi_0 U(\vec z)^\dagger\otimes \Gamma_{\vec z}$ by a
more general quasi-free state. In this subsection, we show that such a
replacement can be accomplished without significant errors.

Let $\Pi$ denote a (particle-number conserving) quasi free state on
$\Ff_<$. It is completely determined by its one-particle density
matrix, which we choose to be given as
\begin{equation}\label{defpi}
\pi = \sum_{|p|< p_c} \pi_p |p\rangle\langle p|\,.
\end{equation}
Here, $|p\rangle \in L^2(\Lambda)$ denotes a plane wave of momentum
$p$. We denote the trace of $\pi$ by $P=\sum_{|p|< p_c} \pi_p$. Let
$\Upsilon_\pi^{\vec z}$ denote the state $\Upsilon_\pi^{\vec z}\equiv
U(\vec z)\Pi U(\vec z)^\dagger \otimes \Gamma_{\vec z}$ on $\Ff$.  We
want to derive an upper bound on the difference
\begin{equation}\label{pipi}
\Tr_\Ff\big[ \dV \big(\Upsilon_\pi^{\vec z}-\Upsilon^{\vec z}\big)\big]= 
\Tr_\Ff\big[ \dV \big(  U(\vec z)\left(\Pi-\Pi_0\right) U(\vec z)^\dagger\otimes
\Gamma_{\vec z}\big)\big]\,. 
\end{equation}
A simple calculation yields
\begin{align}\nonumber
(\ref{pipi}) &= \frac 1{2|\Lambda|} \widehat v(0) \left(  P^2 + 2 P \,
   \Tr_{\Ff_>}\big[ \dN_{\rm s}(\vec z) \Gamma_{\vec z}\big] - 2
\mbox{$\sum_{|k|< p_c}$} \pi_k |z_k|^2\right) \\ \nonumber & \quad
+ \frac 1{2|\Lambda|} \sum_{|k|< p_c,\, |l|< p_c} \widehat
v(k-l) \left[ \pi_k \pi_l + 2 |z_k|^2 \pi_l \right] \\
\nonumber &\quad+
\frac 1{|\Lambda|} \sum_{|k|< p_c,\, |l|\geq p_c} \widehat v(k-l) \pi_k\,
\Tr_{\Ff_>}\big[ \ad{l}\an{l}\Gamma_{\vec z}\big]
\\ &\leq 
\frac{8\pi\varphi}{|\Lambda|} \left( P^2+2 P\, \Tr_{\Ff}\big[
   \dN\Upsilon^{\vec z}\big] \right)\,. \label{pipis}
\end{align}
Here we have used again that $|\widehat v(k)| \leq \widehat v(0) \leq 
8\pi\varphi$. It follows easily from (\ref{assn}) (compare with (\ref{boun}))
that $\Tr_{\Ff}\big[\dN \Upsilon^{\vec z}\big] \leq N(1+ 2/\sqrt{C})$.
Hence we obtain from (\ref{pipis}) that 
\begin{equation}
\Tr_\Ff\big[ \dV  \Upsilon^{\vec z}\big]   \geq
\Tr_\Ff \big[\dV\Upsilon_\pi^{\vec z}\big]  - Z^{(2)}  \,, \label{replv}
\end{equation} 
with
\begin{equation}\label{defz2}
Z^{(2)} =  \frac {8\pi\varphi P^2}{|\Lambda|} + \frac{16\pi P
   \varphi}{|\Lambda|} N \left( 1 + \frac 2{\sqrt{ C}}\right) \,.
\end{equation}
Recall that $C\gg 1$ and $\varphi\gg a$. Hence $Z^{(2)} \ll |\Lambda|a\rho^2$
as long as $\varphi P\ll a N$.

Note that the effect of the replacement of $\Upsilon^{\vec z}$ by
$\Upsilon_\pi^{\vec z}$ on the kinetic energy is 
\begin{equation}\label{notec}
\Tr_\Ff\big[ \dT\Upsilon^{\vec z}\big] = \Tr_\Ff\big[
\dT\Upsilon_\pi^{\vec z}\big] -\sum_{|p|<p_c}
\big(p^2-\mu_0)\pi_p\,.
\end{equation}
We have thus obtained the lower bound 
\begin{align}\nonumber
  F_{\vec z}(\beta) &\geq  \Tr_\Ff\big[ \big( \dT + \dV\big)
  \Upsilon_\pi^{\vec z}\big]
  -\frac 1\beta S(\Upsilon^{\vec z})\\ &\quad -\sum_{|p|<p_c}
\big(p^2-\mu_0)\pi_p + \half  \Tr_{\Ff}\big[ \dK \Upsilon^{\vec
   z}\big] - Z^{(2)}\,. \label{ltx}
\end{align}

\subsection{Dyson Lemma}

Since the interaction potentials in $\dV$ are very short range and
strong (compared with the average kinetic energy per particle), we
cannot directly obtain information on the expectation value of $\dV$
in the state $\Upsilon_\pi^{\vec z}$.  In fact, we cannot even expect
that it yields the desired correction to the free energy, since part
of the interaction energy leading to the second term in (\ref{fmt}) is
actually kinetic energy!  Hence we will first derive a lower bound on
$\dV$ in terms of \lq\lq softer\rq\rq\ and longer ranged potentials,
with the aid of part of the kinetic energy. More precisely, we will
use only the high momentum part of the kinetic energy for this task,
since this is the relevant part contributing to the interaction
energy. The appropriate lemma to achieve this was derived in
\cite{LSSfermi}; part of the idea for such an estimate is already
contained in the paper by Dyson \cite{dyson}. For this reason, we
refer to this estimate as \lq\lq Dyson Lemma\rq\rq.

Our goal is to derive an appropriate lower bound on the
Hamiltonian $\dT + \dV$.  Let $\chi:\R^3\mapsto \R$ be a radial
function, $0\leq \chi(p)\leq 1$, and let
\begin{equation}
h(x)= \frac 1{|\Lambda|} \sum_{p}
\big(1-\chi(p)\big) e^{-ipx} \,.
\end{equation}
We assume that $\chi(p)\to 1$ as $|p|\to\infty$ sufficiently fast
such that
$h\in L^1(\Lambda)\cap L^\infty(\Lambda)$. For some
$L/2>R>R_0$, let
\begin{equation}\label{deffr}
f_R(x)=\sup_{|y|\leq R} | h(x-y) - h(x) |\,,
\end{equation}
and
\begin{equation}\label{defwr}
w_R(x)= \frac 2{\pi^2} f_R(x) \int_{\Lambda} dy\, f_R(y)\,.
\end{equation}
Note that $w_R$ is a periodic function on $\R^3$, with period $L$.

Let $U_R: \R_+\mapsto \R_+$ be a non-negative function that is
supported in the interval $[R_0,R]$, and satisfies $\int_0^\infty dt\,
t^2 U_R(t) \leq 1$.  The following is a simple extension of Lemma~4
(and Corollary~1) in \cite{LSSfermi}. The proof follows closely the
one in \cite{LSSfermi}. For completeness, we present it in the
appendix.

\begin{lem}\label{dlem}
Let $y_1,\dots,y_n$ denote $n$ points in $\Lambda$ and, for
$x\in\Lambda$, let $y_{\rm NN}(x)$ denote the nearest neighbor of $x$
among the points $y_j$, i.e., the $y_k$ minimizing $d(x,y_j)$ among all
$y_j$. We then
have, for any $\eps>0$, 
\begin{equation}\label{dysr}
-\nabla \chi(p)^2 \nabla + \half \sum_{i=1}^n \widetilde v(d(x,y_i)) \geq
  (1-\eps) \widetilde a  U_R(d(x,y_{\rm NN}(x))) - \sum_{i=1}^n  \frac
  {\widetilde a}\eps w_R(x-y_i) \,.
\end{equation}
Here, the operator $-\nabla \chi(p)^2 \nabla$ stands for
$\sum_p p^2 \chi(p)^2 |p\rangle\langle p|$.
\end{lem}

We note that $y_{\rm NN}(x)$ is well defined except on a set of
measure zero. Compared with Lemma~4 in \cite{LSSfermi}, the main
differences are the boundary conditions used, and the fact that we do
not demand a minimal distance between the points $y_i$. In
\cite{LSSfermi}, it was assumed that $d(y_i,y_j)\geq 2R$ for $i\neq
j$, in which case $U_R(d(x,y_{\rm NN}(x)))= \sum_i U_R(d(x,y_j))$. Note
also that only the inequality $\int dt\, t^2 U(t) \leq 1$ is needed
for the estimate, not equality, as stated in \cite{LSSfermi}.

We will use Lemma~\ref{dlem} for a lower bound to the operator
$\dT+\dV$ on $\Ff$. Note that the restriction of this operator to the
sector of $n$ particles is just $\widetilde H_n$, defined in (\ref{ham}). We
write
\begin{equation}\label{hann}
\widetilde H_n = \sum_{j=1}^n \left [ -\Delta_j + \half \sum_{i\neq j}
   \widetilde v(d(x_j,x_i)) \right]\,,
\end{equation} 
and apply the estimate (\ref{dysr}) to each term in square brackets,
for fixed $j$ and fixed positions of the $x_i$, $i\neq j$. We want to
keep a part of the kinetic energy for later use, however.  To this
end, we pick some $0<\kappa<1$, and write
\begin{equation}
p^2 = p^2 (1- (1-\kappa)\chi(p)^2)  + (1-\kappa) p^2 \chi(p)^2 \,.
\end{equation}
We split the kinetic energy in  the
Hamiltonian (\ref{hann}) accordingly, and apply (\ref{dysr}) to the
last part. Using also the positivity of the $\widetilde v$, we thus
obtain, for any subset $J_j \subseteq \{ 1,\dots,j-1,j+1,\dots,n\}$,
\begin{align}\nonumber
  & -\Delta_j + \half \sum_{i\neq j}
   \widetilde v(d(x_j,x_i)) \\  \label{uwone} & \geq -\nabla_j
   (1-(1-\kappa)\chi(p_j)^2) \nabla_j
    \\ & \qquad +(1-\eps)(1-\kappa) \widetilde a U_R(d(x_j,x_{\rm
      NN}^{J_j}(x_j))) - \frac {\widetilde a}\eps
   \sum_{i\in J_j} w_R(x_j-x_i) \,. \nonumber
\end{align}
Here we denoted by $x_{\rm NN}^{J_j}(x_j)$ the nearest neighbor of $x_j$
among the points $x_i$, $i\in J_j$.

Our choice of $J_j$ will depend on the positions $x_i$, $i\neq j$. We
want to choose it in such a way that $d(x_l,x_k)\geq R/5$ if $l\in
J_j$ and $k\in J_j$. Moreover, we want the set to be {\it maximal}, in
the sense that if $l\not\in J_j$, then there exists a $k\in J_j$ such
that $d(x_l,x_k)<R/5$. These properties of $J_j$ will be used in an
essential way in Subsections~\ref{sec29} and~\ref{sec210} below.

There is no unique choice of $J_j$ satisfying these criteria. One way
to construct it is the following. We first pick all $i$ corresponding
to those $x_i$ whose distance to the nearest neighbor (among all the
other $x_k$, $k\neq i,j$) is greater or equal to $R/5$. Secondly,
going through the list $\{x_1,\dots,x_{j-1},x_{j+1},\dots,x_n\}$ one
by one, we add $i$ to the list if $d(x_i,x_j)\geq R/5$ for all $j$
already in the list. This last procedure depends on the ordering of
the $x_i$, and hence the resulting $J_j$ will depend on this ordering.
The resulting interaction potential in (\ref{uwone}) will thus be not
symmetric in the particle coordinates. This is of no importance,
however, since we will take the expectation value of the resulting
operator only in symmetric (bosonic) states anyway.

The set $J_j$ is chosen in order to satisfy the following
requirements. On the one hand, we want the particles keep a certain
minimal distance, $R/5$; this is necessary in order to control the
error terms coming from the potentials $w_R$. We do not have
sufficient control on the two-particle density to control these terms
if all the particle configurations were taken into account.  On the
other hand, we want the balls of radius $R$ centered at the particle
coordinates to be able to overlap sufficiently much, such that the
desired lower bound can be obtained.  We note that we want to derive a
lower bound which is independent of $\vec z$; for certain values of
$\vec z$, however, the system may be far from being homogeneous and
particles may cluster in a relatively small volume. We want to ensure
that there is still sufficient interaction among them.

\subsection{Filling the Holes}\label{holesec}

One defect of Lemma~\ref{dlem} above is that the resulting interaction
potential $U_R$ is supported outside a ball of radius $R_0$, which is
the range of $\widetilde v$. For our estimates in
Subsection~\ref{sec29}, it will be convenient to have a specific
$U_R$ which is, in particular, positive definite and hence should not
have a \lq\lq hole\rq\rq\ at the origin. We will show in this
subsection that one can easily add the missing part to $U_R$, at the
expense of only a small amount of kinetic energy.

We start with the description of our choice of  $U_R$. 
Let $j:\R_+\to \R_+$ denote the \lq\lq hat function\rq\rq 
\begin{equation}\label{defhat}
j(t) = \frac {144}{\pi} \int_{\R^3} dy\, \theta(\half-|y|)
\theta(\half-|y-e t|) 
\end{equation}
for some unit vector $e\in \R^3$. Note that $j$ is supported in the
interval $[0,1]$, and $\int_0^1 dt\, t^2 j(t) = 1$. An explicit
computation yields
\begin{equation}
j(t) = 12 (t+2) [1-t]_+^2\,.
\end{equation}
Our desired interaction potential will be 
$\widetilde U_R(t) = R^{-3} j(t/R)$. We will thus choose $U_R(t)=
\widetilde U_R(t) \theta (t-R_0)$ in (\ref{dysr}).

In the following, it will be convenient to work with $\widetilde U_R$
instead of $U_R$. I.e., we would like the add the missing part
$\widetilde U_R(\,\cdot\,) \theta (R_0-\, \cdot\,)$ to the
interaction. In order to achieve this, we use the following lemma. It
is an easy consequence of the definition of the scattering length,
given in (\ref{defsl}).

\begin{lem}\label{holelem}
Let $y_1,\dots,y_n$ denote $n$ points in $\Lambda$, with
$d(y_i,y_j)\geq R/5$ for $i\neq j$. Let $0\leq \lambda <
\pi/2$, and $R_0< R/10$. Then 
\begin{align}\label{dysrhol}
&-\Delta - \frac{\lambda^2}{R_0^2}  \sum_{i=1}^n \theta(R_0 -  d(x,y_i))
\\ \nonumber & \qquad \geq  - \frac{3 R_0}{(R/10)^3- R_0^3} 
\left(\frac {\tan \lambda}\lambda-1\right) \sum_{i=1}^n \theta(R/10- d(x,y_i))\,.
\end{align}
\end{lem}

\begin{proof}
It suffices to prove that
\begin{align}\nonumber
&\int_{|x|\leq R/10}|\nabla \phi(x)|^2 -  \frac{\lambda^2}{R_0^2}
\int_{|x|\leq R_0} |\phi(x)|^2 \\ & \qquad \geq - \frac{3 R_0}{(R/10)^3- R_0^3} 
\left(\frac {\tan \lambda}\lambda-1\right) \int_{|x|\leq
   R/10}|\phi(x)|^2 \label{sufi}
\end{align}
for any function $\phi \in H^1$. In fact, it is enough to prove
(\ref{sufi}) for radial functions. Note that the scattering length of
the potential $2\lambda^2 R_0^{-2} \theta(R_0-\, \cdot\,)$ is given by
$R_0(1-\lambda^{-1}\tan \lambda)$. Hence, for any $R_0 \leq s\leq R/10$, 
\begin{equation}
\int_{|x|\leq s}|\nabla \phi(|x|)|^2 -  \frac{\lambda^2}{R_0^2}
\int_{|x|\leq R_0} |\phi(|x|)|^2  \geq - 4\pi R_0 
\left(\frac {\tan \lambda}\lambda-1\right)  |\phi(s)|^2\,.
\end{equation}
Eq.~(\ref{sufi}) follows by multiplying this inequality by $s^2$ and
integrating $s$ between $R_0$ and $R/10$.
\end{proof}

Let $\lambda =\pi/4$ for concreteness. Recall that $d(x_i,x_k)\geq R/5$
for $i,k\in J_j$.  Since $\widetilde U_R(t) \leq j(0) /R^3 = 24/R^3$,
and $\widetilde U_R(t) \geq j(1/10)/R^3$ for $t \leq R/10$, this lemma
implies, in particular, that
\begin{align}\label{uhole}
&(\widetilde U_R - U_R)(d(x_j,x_{\rm NN}^{J_j}(x_j)))  \leq - \frac
{24}{\pi^2} \frac{(4 R_0)^2}{R^3}\Delta_j \\ & \quad\qquad +
\frac{18}{(\pi/4)^3}\left(4 - \pi\right) \frac{R_0^3}{(R/10)^3- R_0^3} 
\frac 1{j(1/10)} \widetilde U_R(d(x_j,x_{\rm NN}^{J_j}(x_j)))\,. \nonumber
\end{align}

Define 
\begin{equation}\label{defap}
a' \equiv \widetilde a (1-\eps)(1-\kappa) 
\left( 1- \frac{18}{(\pi/4)^3}\left(4 - \pi\right) \frac{R_0^3}{(R/10)^3- R_0^3} 
\frac 1{j(1/10)}\right)
\end{equation}
and 
\begin{equation}\label{kappapr}
\kappa' \equiv \kappa -  \frac
{24 \widetilde a }{\pi^2} \frac{(4 R_0)^2}{R^3}\,.
\end{equation}
In the following, we will choose $\kappa \gg a R_0^2 / R^3$ and hence,
in particular, $\kappa' > 0$. 
Combining the estimates (\ref{uwone}) and (\ref{uhole}) and applying
them in each sector of particle number $n$, we obtain the inequality
\begin{equation}\label{ltx2t}
\dT + \dV \geq \dT^c + \dW \,, 
\end{equation}
where
\begin{equation}\label{ltx3t}
\dT^c =\sum_p  \eps(p) \ad{p} \an{p} \quad ,
\quad \eps(p) = \kappa' p^2 + (1-\kappa) p^2 (1-\chi(p)^2) - \mu_0\,,
\end{equation}
and $\dW$ is, in each sector with
particle number $n$, given by the (symmetrization of the)
multiplication operator
\begin{equation}\label{defWst}
  \sum_{j=1}^n \biggl[  a' \widetilde U_R(d(x_j,x_{\rm
     NN}^{J_j}(x_j)))  - \frac {\widetilde a}\eps  
\sum_{i\in J_j} w_R(x_j-x_i)\biggl] \,. 
\end{equation}
Note again that the set $J_j$ depends on all the particle coordinates
$x_i$, $i\neq j$.

We now describe our choice of the kinetic energy cutoff $\chi$. 
Let $\nu:\R^3\mapsto \R_+$ be a smooth radial
function with $\nu(p)=0$ for $|p|\leq 1$, $\nu(p)=1$ for $|p|\geq 2$, and
$0\leq \nu(p)\leq 1$ in-between. For some $s\geq R$ we choose
\begin{equation}\label{defchi}
\chi(p)=\nu(s p)\,.
\end{equation}
We will choose $p_c \leq 1/s$ below. This implies, in particular, that
$\eps(p)$, defined in (\ref{ltx3t}) above, is equal to $(1 -
\kappa + \kappa') p^2 - \mu_0$ for $|p|\leq p_c$. Hence (compare with
(\ref{notec}))
\begin{equation}\label{2713}
\Tr_\Ff\big[\dT^c \Upsilon^{\vec z}_\pi\big] = \Tr_\Ff\big[
\dT^c \Upsilon^{\vec z}\big] +\sum_{|p|<p_c}
\big((1-\kappa+\kappa')p^2-\mu_0)\pi_p\,.
\end{equation}

Using the fact that
\begin{equation}
  \Tr_\Ff\big[  \dT^c  \Upsilon^{\vec z}\big]
  -\frac 1\beta S(\Upsilon^{\vec z})
\geq  -\frac 1{\beta} \ln \Tr_{\Ff_>} \exp\big(-\beta
\dT_{\rm s}^c(\vec z)\big)\,,
\end{equation}
we conclude from (\ref{ltx}), (\ref{ltx2t}) and (\ref{2713}) that
\begin{align}\nonumber
F_{\vec z}(\beta) &\geq  -\frac 1{\beta} \ln \Tr_{\Ff_>} \exp\big(-\beta
\dT_{\rm s}^c(\vec z)\big)   + \Tr_\Ff\big[ \dW
   \Upsilon_\pi^{\vec z}\big]+ \half \Tr_{\Ff}\big[ \dK \Upsilon^{\vec
   z}\big] \\ &\quad - (\kappa-\kappa')\sum_{|p|<p_c}
p^2\pi_p  - Z^{(2)} \,. \label{flbF}
\end{align}
Note that the first term on the right side of (\ref{flbF}) can be
computed explicitly. It is given by
\begin{align} \nonumber 
-\frac 1{\beta}\ln \Tr_{\Ff_>} \exp\big(-\beta
\dT_{\rm s}^c(\vec z)\big) & = \sum_{|p|< p_c} \big(
(1-\kappa+\kappa') p^2-\mu_0\big)
|z_p|^2 \\ & \quad + \frac 1\beta  \sum_{|p|\geq p_c}
   \ln \left( 1 - \exp\big(-\beta \eps(p) \right)  \,.
\end{align}
In the following, we will derive a lower bound on the expectation value
of $\dW$ in the state $ \Upsilon_\pi^{\vec z}$.

\subsection{Localization of Relative Entropy}\label{sect:locrel}

Our next task is to give a lower bound on $\Tr_\Ff\big[ \dW
\Upsilon_\pi^{\vec z}\big]$. For that matter, we will show that we can
replace the unknown state $\Gamma_{\vec z}$ in $\Upsilon_\pi^{\vec z}=
U(\vec z)\Pi U(\vec z)^\dagger \otimes \Gamma_{\vec z}$ by the
quasi-free state $\Gamma^0$, which is the Gibbs state for the kinetic
energy $\dT_{\rm s}(\vec z)$. The error in doing so will be controlled
by the upper bound on the relative entropy, Eq.~(\ref{asss}). In order
to do this, we have to obtain a \lq\lq local\rq\rq\ version of this
bound.

Consider the quasi-free state $\Omega_\pi = \Pi \otimes
\Gamma^0$. Its one particle density matrix is given by 
\begin{equation}\label{defell1}
\omega_\pi = \sum_p \frac 1{ e^{\ell(p)}-1}|p\rangle\langle p|\,
\end{equation}
with $\ell(p)= \beta (p^2 - \mu_0)$ for
$|p|\geq p_c$, and $\ell(p) = \ln (1 + 1/{\pi_p})$ for $|p|<p_c$.

Let $\eta:\R^3\mapsto\R$ be a function with the following properties:
\begin{itemize}
\item $\eta\in C^\infty(\R^3)$
\item $\eta(0)=1$, and $\eta(x)=0$ for $|x|\geq 1$
\item $\widehat \eta (p)=\int dx\, \eta(x) e^{-ipx} \geq 0$ for all $p\in\R^3$.
\end{itemize}
An appropriate $\eta$ can, for instance, we obtained by convolving a
smooth function supported in a ball of radius $\half$ with itself.
Given such a function $\eta$, we define $\eta_b(x)=\eta(x/b)$ for some
$b\leq L/2$.  Moreover, with a slight abuse of notation, we define a
one-particle density matrix $\omega_b$ on $\Hh$ by the kernel
\begin{equation}\label{defgd}
\omega_b(x,y)=\omega_\pi(x,y) \eta_b(d(x,y))\,.
\end{equation}
Note that this defines a positive operator, with plane waves as
eigenstates. Note also that $|\omega_b(x,y)|\leq |\omega_\pi(x,y)|$ since
$|\eta_b|\leq 1$. 
We denote by $\Omega_b$ the corresponding (particle
number conserving) quasi-free state
on $\Ff$, and $\Omega_b^{\vec z} = U(\vec z) \Omega_b U(\vec z)^\dagger$. Let
also denote $\rho_\omega = \omega_b(x,x)= \omega_\pi(x,x)$ the one-particle
density of $\Omega_b$ (which is independent of $x$). Abusing the
notation even more, we shall sometimes write
$\omega_b(x,y)=\omega_b(x-y)$ if no confusion can arise.

For $r<L/2$, let $\chi_{r,\xi}(\, \cdot \,) = \theta( r
-d(\,\cdot\,,\xi))$ denote the characteristic function of all ball of
radius $r$ centered at $\xi\in\Lambda$.  The function $\chi_{r,\xi}$
defines a projection on the one-particle space $\Hh=L^2(\Lambda)$, and
hence the Fock space $\Ff$ over $\Hh$ can be thought of as a tensor
product of a Fock space over $\chi_{r,\xi}\Hh$ and a Fock space over
the complement. States on $\Ff$ can thus be restricted to the Fock
space over $\chi_{r,\xi}\Hh$, simply be taking the partial trace over
the other factor. We denote such a restriction of a state $\Gamma$ by
$\Gamma_{\chi_{r,\xi}}$.

For $d(\xi,\zeta)\geq 2r$, $ \chi_{r,\xi}+\chi_{r,\zeta}$ defines a
projection on $\Hh$. Note that since $\omega_{b}(x,y)$ vanishes if
$d(x,y)\geq b$, we have that
\begin{equation}\label{factprop}
\Omega_{b, \chi_{r,\xi}+\chi_{r,\zeta}}=\Omega_{b,
   \chi_{r,\xi}}\otimes \Omega_{b, \chi_{r,\zeta}} 
\end{equation}
if $d(\xi,\zeta)\geq 2r+b$. This follows simply from the fact that the
one particle density matrix of
$\Omega_{b,\chi_{r,\xi}+\chi_{r,\zeta}}$ is given by
$(\chi_{r,\xi}+\chi_{r,\zeta})\omega_b(\chi_{r,\xi}+\chi_{r,\zeta})=
\chi_{r,\xi}\omega_b \chi_{r,\xi}+\chi_{r,\zeta}\omega_b
\chi_{r,\zeta}$. 
The same factorization property (\ref{factprop}) is obviously true with
$\Omega_b$ replaced by $\Omega_b^{\vec z}=U(\vec z)\Omega_b U(\vec
z)^\dagger$ since the unitary $U(\vec z)$ has the same product
structure.  As in \cite[Sect.~5.1]{RSjellium}, we have the following
superadditivity property of the relative entropy.

\begin{lem}\label{lem4}
   Let $X_i$, $0\leq i \leq k$, denote $k$ mutually orthogonal
   projections on $\Hh$. Let $\Omega$ be a state on $\Ff$ which factorizes under
   restrictions as $\Omega_{\sum_i X_i} = \otimes_i \Omega_{X_i}$.
   Then, for any state $\Gamma$,
\begin{equation}
S(\Gamma,\Omega) \geq \sum_i S(\Gamma_{X_i},\Omega_{X_i})\,.
\end{equation}
\end{lem}

We note that the lemma applies, in particular, to a (particle number
conserving) quasi-free state $\Omega$ whose one-particle density
matrix $\omega$ satisfies $X_i\omega X_j=0$ for $i\neq j$. We
emphasize that the factorization property of $\Omega$ is crucial; in
general, the relative entropy need not be superadditive. This is the
reason for introducing the cutoff $b$.

\begin{proof}
Let $X$ denote the projection $X=\sum_i X_i$.  The relative entropy
decreases under restrictions \cite{LRuskai,ohya}, i.e., 
\begin{align}\nonumber
S(\Gamma,\Omega) & \geq S(\Gamma_X,\Omega_X) = S(\Gamma_X,\otimes_i
\Omega_{X_i}) \\ & = \sum_i S(\Gamma_{X_i},\Omega_{X_i}) +  \sum_i
   S(\Gamma_{X_i}) - S(\Gamma_X)\,.\
\end{align}
The last two terms together are positive because of subadditivity of the
von-Neumann entropy.
\end{proof}

We take the $X_i$ to be the multiplication operators by characteristic
functions of balls of radius $r$, separated a distance $2b$. By
averaging over the position of the balls, Lemma~\ref{lem4} implies that, for
any $b\geq 2r$ such that $L/(2b)$ is a positive integer, and for any
state $\Gamma$,
\begin{equation}\label{entes}
S(\Gamma,\Omega_b^{\vec z}) \geq \frac 1{(2b)^3} \int_{\Lambda}d\xi\,
S(\Gamma_{\chi_{r,\xi}},\Omega^{\vec z}_{b,\chi_{r,\xi}})\,.
\end{equation}
We apply this inequality to the state
$\Gamma=\Upsilon_\pi^{\vec z}=U(\vec z)\Pi U(\vec z)^\dagger\otimes
\Gamma_{\vec z}$.

We remark that that restriction of $L/(2b)$ being an integer will be
of no concern to us, since we are interested in the thermodynamic
limit $L\to \infty$, with $b$ independent of $L$.

We can now apply inequality (\ref{sn}) to the right side of
(\ref{entes}). Using the Schwarz inequality for the integration over
$\xi$, we thus obtain
\begin{equation}\label{difs}
\int_{\Lambda}d\xi\,
\|\Upsilon_{\pi,\chi_{r,\xi}}^{\vec
   z}-\Omega_{b,\chi_{r,\xi}}^{\vec z}\|_1 \leq 4 \left(
   b^3 |\Lambda| S(\Upsilon_\pi^{\vec z},\Omega_{b}^{\vec z}) \right)^{1/2}
\end{equation}
for any $r\leq b/2$. Note that $S(\Upsilon_\pi^{\vec
  z},\Omega_{b}^{\vec z}) = S(\Upsilon_\pi, \Omega_b)$ since the
relative entropy is invariant under unitary transformations. Were it
not for the cutoff $b$, we could use (\ref{asss}) to bound the
right side of (\ref{difs}). We will estimate the effect of the cutoff
in Subsection~\ref{relesec}.

\subsection{Interaction Energy, Part 1}\label{sec29}

The next step is to derive a lower bound on $\Tr_\Ff\big[ \dW
\Upsilon_\pi^{\vec z}\big]$. The main input will be the bound
(\ref{difs}) derived in the previous subsection. We split the estimate
into three parts. First, we give a lower bound on the expectation
value of the terms containing $\widetilde U_R$ in (\ref{defWst}). In
the next subsection, we bound the remaining energy containing the
terms $w_R$. Finally, we combine the two estimates in
Subsection~\ref{intesec}. One of the difficulties in our estimates
results form the fact that $\vec z$ is rather arbitrary, and hence the
system can be far from being homogeneous.

From (\ref{defWst}), we can write 
\begin{equation}
\dW  =  \dW_1 - \dW_2 \,,
\end{equation}
where
\begin{equation}\label{defW1}
\dW_1 \equiv \bigoplus_{n=0}^\infty \sum_{j=1}^n  a'\, 
\widetilde U_R(d(x_j,x_{\rm
     NN}^{J_j}(x_j))) 
\end{equation}
and 
\begin{equation}\label{defw2}
\dW_2 = \bigoplus_{n=0}^\infty  \sum_{j=1}^n \sum_{i\in J_j}  \frac{
   \widetilde a }{\eps} w_R(x_j-x_i) \,.
\end{equation}
We start by giving a lower bound on the expectation value of $\dW_1$
in the state $\Upsilon_\pi^{\vec z}$. Recall that $\Upsilon_\pi^{\vec
   z}$ is defined after Eq.~(\ref{defpi}) as $\Upsilon_\pi^{\vec z} =
U(\vec z) \Pi U(\vec z)^\dagger \otimes \Gamma_{\vec z}$.
According to the
decomposition (\ref{defhat}), we can write $\widetilde U_R$ as
\begin{equation}\label{defut}
\widetilde U_R(d(x,y)) = \frac {144}{\pi R^6} \int_{\Lambda} d\xi\,
\theta(R/2-d(\xi,x)) \theta(R/2-d(\xi,y)) \,.
\end{equation}
This gives rise to a corresponding decomposition of $\dW_1$, which we
write as 
\begin{equation}\label{dec1}
\dW_1 =  \frac {144 a'}{\pi R^6} \int_{\Lambda} d\xi\, \dw(\xi)\,.
\end{equation}

For $r>0$, let $n_{r,\xi}$ denote the operator that counts the number
of particles inside a ball of radius $r$ centered at $\xi\in\Lambda$.
It is the second quantization of the multiplication operator
$\chi_{r,\xi}(\,\cdot\,)= \theta (r- d(\xi,\,\cdot\,))$ on
$L^2(\Lambda)$.

We claim that 
\begin{equation}
\dw(\xi) \geq  n_{R/10,\xi} \theta ( n_{R/10,\xi} - 2)
\,.
\end{equation}
This is just the second quantized version of the inequality
\begin{align}\nonumber
& \theta(R/2-d(\xi,x_j)) \theta(R/2-d(\xi,x_{\rm NN}^{J_j}(x_j))) \\ &
\qquad \geq
\theta(R/10-d(\xi,x_j))\left(1- \prod_{i\neq j}
   \theta(d(\xi,x_i)-R/10)\right)\,. \label{topr}
\end{align}
To prove (\ref{topr}), we have to show that whenever $x_j$ and some
$x_k$, $k\neq j$, are in a ball of radius $R/10$ centered at $\xi$,
then $x_{\rm NN}^{J_j}(x_j)$ is in a ball of radius $R/2$ (with the
same center). Assume first that $k\in J_j$. Then $d(x_j,x_{\rm
   NN}^{J_j})\leq d(x_j,x_k)\leq R/5$, whence $d(\xi,x_{\rm
   NN}^{J_j})\leq 3R/10$. If, on the other hand, $k\not\in J_j$, then
there exists an $l\in J_j$ such that $d(x_l,x_k)<R/5$. Hence 
$d(x_j,x_{\rm NN}^{J_j}(x_j))\leq d(x_j,x_l)< 2R/5$, and therefore 
$d(\xi,x_{\rm NN}^{J_j})<R/2$. This proves (\ref{topr}).

Hence, in particular, we have that 
\begin{equation}\label{wxil}
\dw(\xi) \geq \overline \dw(\xi)\equiv  \dw(\xi)\, 
\theta ( 2 - n_{3R/2,\xi} ) +  n_{R/10,\xi} \theta (
n_{R/10,\xi} - 2)  \theta (n_{3R/2,\xi}-3)\,.
\end{equation}
We now claim that
\begin{equation}\label{keye}
\dw(\xi)\, \theta ( 2- n_{3R/2,\xi}) = n_{R/2,\xi}\big( n_{R/2,\xi} -
1\big) \theta (2-n_{3R/2,\xi})\,.\
\end{equation}
This implies, in particular, that the operator $\overline\dw(\xi)$
depends only on the Fock space restricted to a ball of radius $3R/2$
centered at $\xi$. Eq.~(\ref{keye}) follows from the fact that if two
particles with coordinates $x_i$ and $x_j$ are within a ball of radius
$R/2$, and no other particle is in the bigger ball of radius $3R/2$,
then the two particles must be nearest neighbors. Moreover, $j\in J_i$
and $i\in J_j$ by construction.

Note that (\ref{keye}) is a
bounded operator, bounded by $2$. Moreover, since $n_{R/10,\xi}\leq
n_{3R/2,\xi}$, we also see that 
\begin{equation}\label{keyee}
| \overline\dw(\xi) - n_{R/10,\xi}| \leq 2\,.
\end{equation}
Using (\ref{dec1}), (\ref{wxil}) and (\ref{keyee}), we can estimate
\begin{align}\nonumber
   \Tr_\Ff\big[ \dW_1 \Upsilon_\pi^{\vec z}\big] &\geq \frac {144 a'}{\pi R^6}
   \int_{\Lambda} d\xi\, \Tr_\Ff\big[
   \overline\dw(\xi)\Upsilon_\pi^{\vec z}
   \big]\\ \nonumber 
&\geq \frac {144 a'}{\pi R^6} \int_{\Lambda} d\xi\,
   \Tr_\Ff\big[ \overline\dw(\xi)\Omega_b^{\vec z} +
   n_{R/10,\xi}\big(\Upsilon^{\vec z}_{\pi} -
   \Omega_b^{\vec z}\big)
\big] \\ & \quad  - 2 \frac {144 a'}{\pi R^6}
   \int_{\Lambda} d\xi\, \| \Upsilon^{\vec z}_{\pi,\chi_{3R/2,\xi}} -
   \Omega^{\vec z}_{b,\chi_{3R/2,\xi}}\|_1 \,. \label{ott}
\end{align}
Here we have also used that $\overline\dw(\xi)$ acts non-trivially
only on the Fock space over $\chi_{3R/2,\xi}\Hh$. 
Note that the integral over the second term on the right side of
(\ref{ott}) is equal to 
\begin{equation}
  \int_{\Lambda}
d\xi\, \Tr_{\Ff}\big[
n_{R/10,\xi}\big(\Upsilon_{\pi}^{\vec z} -
\Omega_b^{\vec z}\big)\big] = 
\frac{4\pi}{3} \left(\frac R{10}\right)^3 \Tr_\Ff\big[
\dN\big(\Upsilon_\pi^{\vec z}-\Omega_b^{\vec z}\big)\big]\,.\label{onon}
\end{equation}
Moreover, for the last term in (\ref{ott}), we can use (\ref{difs}) to
estimate
\begin{equation}\label{toto}
  \int_{\Lambda} d\xi\, \| \Upsilon^{\vec z}_{\pi,\chi_{3R/2,\xi}} -
   \Omega^{\vec z}_{b,\chi_{3R/2,\xi}}\|_1 \leq 4 \left(
   b^3 |\Lambda| S(\Upsilon_\pi^{\vec z},\Omega_{b}^{\vec z}) \right)^{1/2}
\end{equation}
as long as $3 R \leq b $.

We proceed with a lower bound on $ \Tr_\Ff\big[
\overline\dw(\xi)\Omega_b^{\vec z}\big]$. In fact, we will derive two
different lower bounds on this expression. First, neglecting the last
term in (\ref{wxil}) and using (\ref{keye}), 
\begin{align}\nonumber
  \Tr_\Ff\big[ \overline\dw(\xi)\Omega_b^{\vec z}\big] \geq & \biggl[
  \Tr_\Ff\big[ n_{R/2,\xi}\big(n_{R/2,\xi}-1\big)\Omega_b^{\vec
    z}\big] \\ & - \Tr_\Ff \big[
  n_{3R/2,\xi}\big(n_{3R/2,\xi}-1\big)\big(n_{3R/2,\xi}-2\big)\Omega_b^{\vec
    z}\big]\biggl]_+\,. \label{3pe}
\end{align} 
Since $\Omega_b^{\vec z}$ is a combination of a coherent and
quasi-free state, the last expression in (\ref{3pe}) is, in fact, easy
to estimate. Let $\Phi_{\vec z}$ denote the one-particle state
$|\Phi_{\vec z}\rangle = \sum_{|p|<p_c} z_p |p\rangle$.  We have
\begin{align}\nonumber
  &\Tr_\Ff \big[
  n_{3R/2,\xi}\big(n_{3R/2,\xi}-1\big)\big(n_{3R/2,\xi}-2\big)\Omega_b^{\vec
    z}\big] \\ \nonumber & = \left( \Tr_\Ff\big[ n_{3R/2,\xi}
    \Omega_b^{\vec z}\big] \right)^3 +2\, \tr\, (\chi_{3R/2,\xi}
  \omega_b)^3 + 6 \langle\Phi_{\vec z}|
  (\chi_{3R/2,\xi} \omega_b \chi_{3R/2,\xi})^2 |\Phi_{\vec z}\rangle \\
  \nonumber & \quad + 3 \left( \Tr_\Ff\big[ n_{3R/2,\xi}
    \Omega_b^{\vec z} \big]\right) \left( 2 \langle \Phi_{\vec z}|
    \chi_{3R/2,\xi} \omega_b \chi_{3R/2,\xi}|\Phi_{\vec z}\rangle +
    \tr\, ( \chi_{3R/2,\xi} \omega_b)^2 \right) \\ & \leq 6 \left(
    \Tr_\Ff\big[ n_{3R/2,\xi} \Omega_b^{\vec z}\big] \right)^3\,.
\label{cq3}
\end{align}
(Here, we use the symbol $\tr$ to denote the trace over the
one-particle space $L^2(\Lambda)$, while $\Tr$ is reserved for the
trace over the Fock space.)

A different lower bound can be obtained using 
\begin{equation}\label{difl}
\Tr_\Ff \big[ \overline\dw(\xi)\Omega_b^{\vec z}\big] 
\geq \Tr_\Ff \big[ n_{R/10,\xi} \theta (
n_{R/10,\xi} - 2) \Omega_b^{\vec z}\big]\,.
\end{equation}
Eq.~(\ref{difl}) follows easily from (\ref{wxil}) and (\ref{keye}).
The latter trace is non-trivial only over the Fock space over
$\chi_{R/10,\xi}\Hh$. Denoting by $\Pi_0^\Ff$ the vacuum on $\Ff$, we
claim that
\begin{equation}\label{vacc}
\Omega_{b,\chi_{R/10,\xi}} \geq e^{- 4\pi (R/10)^3 \rho_\omega/3} 
\, \Pi^\Ff_{0,\chi_{R/10,\xi}}\,,
\end{equation}
which implies that 
\begin{equation}\label{impl19}
\Omega^{\vec z}_{b,\chi_{R/10,\xi}} \geq e^{- 4\pi (R/10)^3
  \rho_\omega/3} 
\,  \big( U(\vec z)^\dagger \Pi_0^\Ff
U(\vec z)\big)_{\chi_{R/10,\xi}}\,.
\end{equation}
Eq.~(\ref{vacc}) follows from the fact that $\Omega_{
   b,\chi_{R/10,\xi}}$ is a particle-number conserving quasi-free state, whose
vacuum part is given by 
\begin{align}\nonumber
\exp\big( - \tr\, \ln (1+
\chi_{R/10,\xi}\omega_b  \chi_{R/10,\xi})\big) & \geq \exp\big( - \tr\,
\chi_{R/10,\xi}\omega_b  \chi_{R/10,\xi})\big) \\ & =  \exp\big(- 4\pi
(R/10)^3 \rho_\omega\big/3)\,.
\end{align}

Eq.~(\ref{impl19}) implies, in particular, that 
\begin{equation}
(\ref{difl}) \geq e^{- 4\pi (R/10)^3 \rho_\omega/3}\,
\Tr_\Ff \big[  n_{R/10,\xi} \theta (
n_{R/10,\xi} - 2) U(\vec z)^\dagger \Pi_0^\Ff U(\vec z)\big]\,.
\end{equation}
The state $ U(\vec z)^\dagger \Pi_0^\Ff U(\vec z)$ is a coherent state
on $\Ff$. Its restriction to the Fock space over $\chi_{R/10,\xi}\Hh$
is again a coherent state. In every sector of particle number $n$, it
is given by the projection onto the $n$-fold tensor product of the
wave function $\chi_{R/10,\xi}\Phi_{\vec z}$, appropriately
normalized. Therefore
\begin{align}\nonumber
  &\Tr_\Ff \big[ n_{R/10,\xi} \theta (
  n_{R/10,\xi} - 2) U(\vec z)^\dagger \Pi_0^\Ff U(\vec z)\big] \\
  &= e^{ - \langle \Phi_{\vec z} | \chi_{R/10,\xi}|\Phi_{\vec
      z}\rangle } \sum_{n\geq 2} n \frac { \langle \Phi_{\vec z} |
    \chi_{R/10,\xi}|\Phi_{\vec z}\rangle^n } {n!} \\ \nonumber & =
  \langle \Phi_{\vec z} | \chi_{R/10,\xi}|\Phi_{\vec z}\rangle \left(
    1 - e^{-\langle \Phi_{\vec z} | \chi_{R/10,\xi}|\Phi_{\vec
        z}\rangle } \right) \geq \frac{ \langle \Phi_{\vec z} |
    \chi_{R/10,\xi}|\Phi_{\vec z}\rangle^2 }{1 + \langle \Phi_{\vec z}
    | \chi_{R/10,\xi}|\Phi_{\vec z}\rangle} \,.
\end{align}
The last inequality follows from the elementary estimate
$x(1-e^{-x})\geq x^2/(1+x)$ for $x\geq 0$.

Summarizing the results of this subsection, we have shown that, for any
$0\leq \lambda\leq 1$, 
\begin{align}\nonumber
  & \Tr_\Ff\big[ \dW_1 \Upsilon_\pi^{\vec z}\big] \\ \nonumber & \geq
  \frac{24}{125} \frac{a'}{R^3} \Tr_\Ff\big[ \dN \big(
  \Upsilon_\pi^{\vec z}-\Omega_b^{\vec z}\big)\big] - 144 \frac{8}\pi
  \frac {a'}{R^6} \left( b^3 |\Lambda| S(\Upsilon_\pi^{\vec
      z},\Omega_b^{\vec z})\right)^{1/2}\\ \nonumber & \quad + \lambda
  \frac {144}\pi \frac {a'}{R^6} \int_\Lambda d\xi \left[
    \Tr_\Ff\big[n_{R/2,\xi}(n_{R/2,\xi}-1)\Omega_b^{\vec z}\big] - 6
    \left(\Tr_\Ff\big[n_{3R/2,\xi}\Omega_b^{\vec z}\big]\right)^3
  \right]_+ \\ & \quad + (1-\lambda) \frac {144}\pi \frac
  {a'}{R^6}e^{- 4\pi(R/10)^3 \rho_\omega/3} \int_\Lambda d\xi\, \frac{
    \langle \Phi_{\vec z} | \chi_{R/10,\xi}|\Phi_{\vec z}\rangle^2 }{1
    + \langle \Phi_{\vec z} | \chi_{R/10,\xi}|\Phi_{\vec z}\rangle}
  \,. \label{sumw1}
\end{align}
The choice of $\lambda$ will depend on the function $|\Phi_{\vec z}|$.
If it is approximately a constant (in a sense to be made precise in
Subsection~\ref{intesec}), we will take $\lambda=1$, otherwise we
choose $\lambda=0$.

\subsection{Interaction Energy, Part 2}\label{sec210}

Next we are going to give an upper bound on the expectation value of
$\dW_2$, defined in (\ref{defw2}). To start, we claim that there
exists a smooth function $g$ of rapid decay (faster than any
polynomial) such that
\begin{equation}\label{wg}
w_R(x-y) \leq  \frac{R^2}{s^5} g (d(x,y)/s)\,.
\end{equation}
Although $w_R$ depends on the box size $L$, $g$ can be chosen
independent of $L$ for large $L$. This follows immediately from the
following considerations. First of all, we have, from the definition
(\ref{deffr}) of $f_R$ and because of $R\leq s$,
\begin{equation}\label{fffh}
f_R(x) \leq R \sup_{d(x,y)\leq s} |\nabla h(y)|\,.
\end{equation}
Recall that $h(x)= |\Lambda|^{-1} \sum_p (1-\vu(sp))e^{-ipx}$, where
$1-\vu$ is a smooth function supported in a ball of radius $2$. We
need the following elementary lemma.

\begin{lem}
Let $o:\R^3\to \C$ be a smooth function, supported in a cube of side length
$4$, and let $u(x)=|\Lambda|^{-1} \sum_p o(s p) e^{-ipx}$. Then, for any
non-negative integer $n$, 
\begin{equation}
|u(x)| \leq  \left(\frac{s}{16\, d(x,0)}\right)^{2n} \|
(-\Delta)^n o\|_\infty \left(\frac 2{\pi s} +  2 \frac{n+1}L\right)^3\,.
\end{equation}
\end{lem}

Here, $\Delta$ denotes the Laplacian on $\R^3$, not on $\Lambda$.

\begin{proof}
Introducing coordinates $x=(x_1,x_2,x_3)$, we can write
\begin{equation}\label{cubb}
u(x) \left( 2 L^2\sum_{i=1}^3 (1-\cos(2\pi x_i/L))\right)^n =
|\Lambda|^{-1} \sum_p (-\Delta_{\rm d})^n o(sp) e^{-ipx}\,.
\end{equation}
Here, $\Delta_{\rm d}$ denotes the discrete Laplacian in momentum
space, which acts as $L^{-2} (-\Delta_d)f(p) = 8 f(p) -
\sum_{|e|=1}f(p+2\pi e/L)$. It is easy to see that the function
$(-\Delta_d)^n f$ is bounded by $\|(-\Delta)^n f\|_\infty$. Moreover,
if $f$ has support in a cube of side length $\ell$, then $(-\Delta_{\rm
   d})^n f$ is supported in a cube of length $\ell + 4\pi n/L$. This implies
that
\begin{equation}
|(\ref{cubb})| \leq s^{2n} \|(-\Delta)^n o\|_\infty
\left(\frac 2{\pi s} +  2 \frac{n+1}L\right)^3 \,.
\end{equation}
On the other hand, note that $1-\cos(2\pi x_i/L)\geq 8 L^{-2}
\min_{k\in \Z}|x_i-kL|^2$, and hence
\begin{equation}
  2 L^2 \sum_{i=1}^3 (1-\cos(2\pi x_i/L)) \geq 16\, d(x,0)^2\,.
\end{equation}
This proves the lemma.
\end{proof}

Applying the lemma to the function $\nabla h$ in (\ref{fffh}), and using the
definition (\ref{defwr}) of $w_R$, we immediately
conclude (\ref{wg}).

We now decompose the function $g$ into an integral over characteristic
functions of balls. Such decompositions have been studied in detail in
\cite{hsdec}. Recall that $j$ is defined in (\ref{defhat}). According to
\cite[Thm.~1]{hsdec}, we can write
\begin{equation}
g(t) = \int_{0}^\infty dr\, m(r) j(t/r) \,,
\end{equation}
where 
\begin{equation}
m(r) = \frac 1{72} r \left( g''(r) - r g'''(r)\right)\,.
\end{equation}
Note that $m$ is a smooth function of rapid decay. 
Since $j$ is monotone decreasing, we can estimate
\begin{equation}
g(t) \leq    j(t)\int_0^1 dr\, | m(r)| + \int_1^\infty dr\, |m(r)|\,
j(t/r)\,.
\end{equation}
This estimate, together with (\ref{wg}), implies that 
\begin{align}\nonumber
   \dW_2 & \leq \frac{144}{\pi} \frac{\widetilde aR^2}{\eps s^8}
   \int_s^\infty dr\, \left( \delta(r-s) \mbox{$ \int_0^1 dt\, |
       m(t)|$} + s^{-1}|m(r/s)| \right) \\ &\quad \qquad\qquad\qquad
   \times \int_\Lambda d\xi\, \bigoplus_{n=0}^\infty \sum_{j=1}^n
   \sum_{i\in J_j}\chi_{r/2,\xi}(x_j)\chi_{r/2,\xi}(x_i)\,.\label{wv}
\end{align}

Let $v_r(\xi)$ denote the integrand in the last line in (\ref{wv}).
Because $d(x_i,x_k)\geq R/5$ for $i,k\in J_j$, the number of $x_i$
inside a ball of radius $r/2$ is bounded from above by $(1+5 r/R)^3$.
Thus we have
\begin{equation}\label{21011}
v_r(\xi) \leq n_{r/2,\xi}(1+5 r/R)^3\,.
\end{equation}
Moreover, we trivially have that $v_r(\xi)\leq
n_{r/2,\xi}(n_{r/2,\xi}-1)$. By combining these two bounds, we obtain
\begin{equation}
v_r(\xi) \leq f(n_{r/2,\xi})\,, 
\end{equation}
where
\begin{equation}
f(n) = n \min\{ n-1, (1+5 r/R)^3 \}\,.
\end{equation}
Proceeding similarly to (\ref{ott}), using that $|f(n)- n
(1+5r/R)^3|\leq (1+ (1+5r/R)^3)/4$,  we can estimate
\begin{align}\nonumber
  \Tr_{\Ff}\big[ v_r(\xi) \Upsilon_\pi^{\vec z}\big] & \leq
  \Tr_{\Ff}\big[ f(n_{r/2,\xi}) \Upsilon_\pi^{\vec z}\big] \\
  \nonumber &\leq \Tr_{\Ff}\big[ f(n_{r/2,\xi}) \Omega_b^{\vec z}\big]
  + (1+5 r/R)^3 \Tr_{\Ff}\big[ n_{r/2,\xi} \big( \Upsilon_\pi^{\vec z}
  -\Omega_b^{\vec z}\big)\big] \\ &\quad +\mbox{$\frac 14$}\left( 1 +
    (1+5 r/R)^3\right)^2 \| \label{xxo} \Upsilon^{\vec
    z}_{\pi,\chi_{r/2,\xi}}-\Omega^{\vec z}_{b,\chi_{r/2,\xi}}\|_1\,.
\end{align}
When integrating over $\xi$, the last two terms can be handled in the
same way as in the previous subsection, see Eqs.~(\ref{onon}) and
(\ref{toto}). We have to assume that $r\leq b$, however. For the first
term on the right side of (\ref{xxo}), we estimate
\begin{align}\nonumber
&\Tr_{\Ff}\big[ f(n_{r/2,\xi}) \Omega_b^{\vec z}\big]  \\ & \leq \min\left\{
   \Tr_{\Ff}\big[ n_{r/2,\xi}(n_{r/2,\xi}-1) \Omega_b^{\vec
     z}\big]\,,\,(1+5 r/R)^3 \Tr_\Ff\big[ n_{r/2,\xi}\Omega_b^{\vec
     z}\big]\right\}\,.
\end{align} 
Similarly to (\ref{cq3}),
\begin{equation}
  \Tr_{\Ff}\big[ n_{r/2,\xi}(n_{r/2,\xi}-1) \Omega_b^{\vec
     z}\big] \leq 2 \left( \Tr_\Ff\big[ n_{r/2,\xi}\Omega_b^{\vec
     z}\big] \right)^2 \,,
\end{equation}
and hence
\begin{equation}
\Tr_{\Ff}\big[ f(n_{r/2,\xi}) \Omega_b^{\vec z}\big] \leq 
\frac { 4 \left(\Tr_\Ff\big[ n_{r/2,\xi}\Omega_b^{\vec
     z}\big]\right)^2}{1+ 2\, \Tr_\Ff\big[ n_{r/2,\xi}\Omega_b^{\vec
     z}\big]/\left(1 + 5 r/R\right)^3}\,.
\end{equation}
Moreover,
\begin{equation}
  \Tr_\Ff\big[ n_{r/2,\xi}\Omega_b^{\vec
     z}\big] = \frac \pi 6 r^3 \rho_\omega + \langle \Phi_{\vec z}|
   \chi_{r/2,\xi}|\Phi_{\vec z}\rangle\,.
\end{equation}
Using convexity of the function $x\mapsto x^2/(1+x)$,
  we obtain the bound
\begin{equation}\label{fbo}
\Tr_{\Ff}\big[ f(n_{r/2,\xi}) \Omega_b^{\vec z}\big]
  \leq 8 \left(
   \frac \pi 6 r^3 \rho_\omega \right)^2 + \frac { 8  \langle \Phi_{\vec z}|
   \chi_{r/2,\xi}|\Phi_{\vec z}\rangle^2}{1+  4\langle \Phi_{\vec z}|
   \chi_{r/2,\xi}|\Phi_{\vec z}\rangle/(1+5 r/R)^3}
\,.
\end{equation}
We use (\ref{fbo}) in (\ref{xxo}) and integrate  over $\xi$. We obtain
(assuming $r\leq b$, as mentioned above) 
\begin{align}\nonumber
  \int_\Lambda d\xi\, \Tr_{\Ff}\big[ v_r(\xi) \Upsilon_\pi^{\vec
    z}\big] & \leq (1+5 r/R)^3 \frac \pi 6 r^3 \Tr_\Ff\big[\dN
  \big(\Upsilon_\pi^{\vec z} - \Omega_b^{\vec z}\big)\big] + 8
  |\Lambda| \left( \frac \pi 6 r^3 \rho_\omega \right)^2 \\ \nonumber
  &\quad + \left( 1 + (1+5 r/R)^3\right)^2 \left(b^3 |\Lambda|
    S(\Upsilon_\pi^{\vec z},\Omega_{b}^{\vec z})\right)^{1/2} \\ &
  \quad + \int_\Lambda d\xi\, \frac { 8 \langle \Phi_{\vec z}|
    \chi_{r/2,\xi}|\Phi_{\vec z}\rangle^2}{1+ 4\langle \Phi_{\vec z}|
    \chi_{r/2,\xi}|\Phi_{\vec z}\rangle/(1+5 r/R)^3} \label{21020} \,.
\end{align}
In order to be able to compare the last term with the last term in
(\ref{sumw1}), we note that
\begin{equation}\label{smea}
\chi_{r/2,\xi} \leq \left( 1+ \frac{5r}R\right)^3  \dashint_{|a|\leq
   r/2+R/10} da \, \chi_{R/10,\xi+a}\,.
\end{equation}
Here, we denote by $\dashint$ the normalized integral, i.e., we divide
by the volume of the ball of radius $r/2+R/10$. Using monotonicity and
convexity of the map $x\mapsto x^2/(1+x)$, we thus have the upper
bound
\begin{align}\nonumber
  & \frac { \langle \Phi_{\vec z}| \chi_{r/2,\xi}|\Phi_{\vec
      z}\rangle^2}{1+ 4\langle \Phi_{\vec z}|
    \chi_{r/2,\xi}|\Phi_{\vec z}\rangle/(1+5 r/R)^3} \\ &\leq \left(
    1+ \frac{5r}R\right)^6 \dashint_{|a|\leq r/2+R/10} da\, \frac {
    \langle \Phi_{\vec z}| \chi_{R/10,\xi+a}|\Phi_{\vec
      z}\rangle^2}{1+ 4 \langle \Phi_{\vec z}|
    \chi_{R/10,\xi+a}|\Phi_{\vec z}\rangle}\,. \label{red}
\end{align}
After integration over $\xi$, the right side of (\ref{red}) simply becomes
\begin{equation}
\left( 1+ \frac{5r}R\right)^6 \int_{\Lambda} d\xi \,  
\frac {  \langle \Phi_{\vec z}|
   \chi_{R/10,\xi}|\Phi_{\vec z}\rangle^2}{1+ 4 \langle \Phi_{\vec z}|
   \chi_{R/10,\xi}|\Phi_{\vec z}\rangle} \leq 
\left(\frac{6r}R\right)^6 \int_{\Lambda} d\xi \,  
\frac {  \langle \Phi_{\vec z}|
   \chi_{R/10,\xi}|\Phi_{\vec z}\rangle^2}
{1+  \langle \Phi_{\vec z}|
   \chi_{R/10,\xi}|\Phi_{\vec z}\rangle} 
\,.
\end{equation}
Here we have used the fact that $r\geq s \geq R$ for the relevant
values of $r$.

As noted above, the estimates leading to (\ref{21020}) are only valid
for $r\leq b$. To bound the expectation value of $\dW_2$ in
(\ref{wv}) we have to consider all $r\geq s$, however. 
For $r\geq b$, we use (\ref{21011}) to obtain the simple estimate
\begin{align}\nonumber
  \int_\Lambda d\xi\, \Tr_\Ff\big[ v_r(\xi) \Upsilon_\pi^{\vec z}\big]
  & \leq \left( 1+ \frac{5r}R\right)^3 \int_\Lambda d\xi\,
  \Tr_\Ff\big[ n_{r/2,\xi} \Upsilon_\pi^{\vec z}\big] \\ & \leq \left(
    \frac{6r}R\right)^3 \frac \pi 6 r^3 \Tr_\Ff\big[\dN
  \Upsilon_\pi^{\vec z}\big]\,.
\end{align}
The contribution of $r\geq b$ to the integral in (\ref{wv}) is thus
bounded from above by 
\begin{align} \nonumber 
& \frac 1s \int_b^\infty dr\, |m(r/s)| \int_\Lambda d\xi\, \Tr_\Ff\big[
v_r(\xi) \Upsilon_\pi^{\vec z}\big]\\  & \quad \leq \frac \pi 6 s^3
\left(\frac{6s}{R}\right)^3  \Tr_\Ff\big[\dN
\Upsilon_\pi^{\vec z}\big]\, \int_{b/s}^\infty dr\, r^6 |m(r)| \,.
\end{align}
Since $|m|$ is a function that decays faster than any polynomial, the
last integral is bounded above by any power of the (small) parameter $s/b$.

Let $c$ denote the constant 
\begin{equation}
c=   \int_0^1 dr\, |m(r)| +  \int_1^\infty dr\, r^6
     |m(r)| \,.
\end{equation}
To summarize, we have derived the upper bound
\begin{align}\nonumber
  \Tr_\Ff\big[\dW_2 \Upsilon_\pi^{\vec z}\big] & \leq 6^3 \frac { 24
    \widetilde a}{\eps R s^2}c \, \Tr_\Ff\big[
  \dN\big(\Upsilon_\pi^{\vec z}-\Omega_b^{\vec z}\big)\big]+ 32\pi
  |\Lambda| \widetilde a\rho_\omega^2 c \frac{R^2}{\eps s^2} \\ 
  \nonumber & \quad + \frac{144}\pi (1+6^3)^2 \frac{ \widetilde a
  }{\eps R^4 s^2}c \left( b^3 |\Lambda|
    S(\Upsilon_\pi^{\vec z},\Omega_b^{\vec z})\right)^{1/2} \\
  \nonumber & \quad + 6^3 \frac{24 \widetilde a}{\eps R s^2}
  \Tr_\Ff\big[\dN \Omega_b^{\vec z}\big] \, \int_{b/s}^\infty dr\, r^6
  |m(r)| \\ &\quad + 8 \frac{144}\pi \frac{ 6^6 \widetilde a}{\eps s^2
    R^4} c \int_{\Lambda} d\xi \, \frac { \langle \Phi_{\vec z}|
    \chi_{R/10,\xi}|\Phi_{\vec z}\rangle^2}{1+ \langle \Phi_{\vec z}|
    \chi_{R/10,\xi}|\Phi_{\vec z}\rangle}\,.\label{sumw2}
\end{align}

\subsection{Interaction Energy, Part 3}\label{intesec}

We now put the bounds of the previous two subsections together in
order to obtain our final lower bound on
$\Tr_\Ff[\dW\Upsilon_\pi^{\vec z}]$.  We will distinguish two cases,
depending on the value of a certain function of $|\Phi_{\vec z}|$,
given in (\ref{asph}) below.

Assume first that 
\begin{equation}\label{asph}
  \int_{\Lambda} d\xi \,  \frac {  \langle \Phi_{\vec z}|
   \chi_{R/10,\xi}|\Phi_{\vec z}\rangle^2}{1+  \langle \Phi_{\vec z}|
   \chi_{R/10,\xi}|\Phi_{\vec z}\rangle} \geq \frac{\pi^2}{18}|\Lambda|
(R^3 \rho)^2 \,.
\end{equation}
This condition means, essentially, that $|\Phi_{\vec z}|$ is far from
being a constant.  In this case, we choose $\lambda=0$ in
(\ref{sumw1}), and find that the contribution of the last terms in
(\ref{sumw1}) and (\ref{sumw2}), respectively, is bounded from below
by
\begin{equation}
8\pi |\Lambda| \widetilde a \rho^2 \left(\frac{a'}{\widetilde a} 
- \frac {4\pi}3
   \left(\frac R {10}\right)^3\rho_\omega 
- 8 c \frac{6^6 R^2}{\eps s^2} \right)\,.
\end{equation}

Next, consider the case when (\ref{asph}) is false. In this case,
using (\ref{smea}) for $r=3R$, as well as convexity of $x\mapsto
x^2/(1+x)$, we find that
\begin{equation}\label{c2as}
  \int_{\Lambda} d\xi \,  \frac {  \langle \Phi_{\vec z}|
   \chi_{3R/2,\xi}|\Phi_{\vec z}\rangle^2}{1+ 16^{-3} 
\langle \Phi_{\vec z}|
   \chi_{3R/2,\xi}|\Phi_{\vec z}\rangle} 
\leq 16^6 \frac{\pi^2}{18}|\Lambda| (R^3 \rho)^2\,.
\end{equation}
Pick some $D>0$, and let $\Bb\subset \Lambda$ denote the set
\begin{equation}\label{defbb}
\Bb = \left\{ \xi \in \Lambda\,:\,  \langle \Phi_{\vec z}|
   \chi_{3R/2,\xi}|\Phi_{\vec z}\rangle \geq 16^3 D R^3 \rho\right\}\,.
\end{equation}
Using (\ref{c2as}), as well as monotonicity of $x\mapsto x/(1+x)$, we
find that
\begin{equation}
\int_{\Bb} d\xi\,
  \langle \Phi_{\vec z}|
   \chi_{3R/2,\xi}|\Phi_{\vec z}\rangle \leq \frac{16^3}D
   \frac{\pi^2}{18} |\Lambda| R^3\rho
\left( 1 +  D R^3\rho\right)\,.\label{pbe}
\end{equation}
Similarly, we have the estimate
\begin{equation}
|\Bb| \leq |\Lambda| \frac{1}{D^2}
   \frac{\pi^2}{18} \left( 1 +  D R^3\rho\right)\,.\label{be}
\end{equation}
We choose $\lambda = 1$ in (\ref{sumw1}) and estimate the relevant
term from below by 
\begin{align}\nonumber
  & \int_\Lambda d\xi \left[
    \Tr_\Ff\big[n_{R/2,\xi}(n_{R/2,\xi}-1)\Omega_b^{\vec z}\big] - 6
    \left(\Tr_\Ff\big[n_{3R/2,\xi}\Omega_b^{\vec z}\big]\right)^3
  \right]_+ \\ & \geq \int_{\Lambda \setminus \Bb} d\xi \left(
    \Tr_\Ff\big[n_{R/2,\xi}(n_{R/2,\xi}-1)\Omega_b^{\vec z}\big] - 6
    \left(\Tr_\Ff\big[n_{3R/2,\xi}\Omega_b^{\vec z}\big]\right)^3
  \right)\,. \label{ffis}
\end{align}
Using $\Tr_{\Ff}\big[n_{3R/2,\xi} \Omega_b^{\vec z}\big] = 9\pi R^3
\rho_\omega/2 + \langle \Phi_{\vec z} | \chi_{3R/2,\xi} | \Phi_{\vec
  z}\rangle$, the definition of $\Bb$ in (\ref{defbb}) and convexity
of $x\mapsto x^3$, we can bound the last term as
\begin{equation}
\int_{\Lambda\setminus\Bb}d\xi 
\left(\Tr_\Ff\big[n_{3R/2,\xi}\Omega_b^{\vec z}\big]\right)^3
\leq 4 |\Lambda| \left( \frac {9\pi} 2 R^3\rho_\omega\right)^3 +
18 \pi |\vec z|^2 R^3
\left( 16^3 D R^3\rho\right)^2\,. 
\end{equation}

We now investigate the first term on the right side of (\ref{ffis}). A
simple calculation shows that
\begin{align}\nonumber
  &\Tr_{\Ff}\big[ n_{R/2,\xi}\big( n_{R/2,\xi}-1\big)\Omega_b^{\vec
    z}\big] \\ \nonumber &= \Tr_{\Ff}\big[ n_{R/2,\xi}\big(
  n_{R/2,\xi}-1\big)\Omega_b\big] + 2 \langle\Phi_{\vec
    z}|\chi_{R/2,\xi}\omega_b\chi_{R/2,\xi}|\Phi_{\vec z}\rangle \\ &
  \quad + \frac \pi 3 R^3\rho_\omega\, \langle\Phi_{\vec
    z}|\chi_{R/2,\xi}|\Phi_{\vec z}\rangle + \langle\Phi_{\vec
    z}|\chi_{R/2,\xi}|\Phi_{\vec z}\rangle^2\,. \label{nonu}
\end{align}
Here, we have used again the translation invariance of $\Omega_b$.
Note that this invariance also implies that the first term on the
right side of (\ref{nonu}) is independent of $\xi$. Since $\Omega_b$
is a quasi free state, it can be rewritten in terms of the
one-particle density matrix $\omega_b$ as
\begin{equation}
  \Tr_{\Ff}\big[ n_{R/2,\xi}\big(
n_{R/2,\xi}-1\big)\Omega_b\big] = \left(\tr\,
   \chi_{R/2,\xi}\omega_b\right)^2 + 
\tr\, \chi_{R/2,\xi}\omega_b\chi_{R/2,\xi}\omega_b\,.
\end{equation}
The first term is just $(\pi R^3 \rho_\omega/6)^2$, and the second is
bounded from above by this expression. Therefore, 
\begin{equation}
\int_{\Bb} d\xi\, \Tr_{\Ff}\big[ n_{R/2,\xi}\big(
n_{R/2,\xi}-1\big)\Omega_b\big] \leq 2|\Bb|  
\left( \frac \pi 6 R^3 \rho_\omega\right)^2
\,.
\end{equation}
Note also that $\langle\Phi_{\vec
   z}|\chi_{R/2,\xi}\omega_b\chi_{R/2,\xi}|\Phi_{\vec z}\rangle \leq
\tr \chi_{R/2,\xi}\omega_b\, \langle\Phi_{\vec z}|\chi_{R/2,\xi}|\Phi_{\vec
   z}\rangle $, and thus
\begin{align}\nonumber
& \int_{\Bb} d\xi\, \left( 2 \langle\Phi_{\vec
   z}|\chi_{R/2,\xi}\omega_b\chi_{R/2,\xi}|\Phi_{\vec z}\rangle +
\frac \pi 3  R^3\rho_\omega\,
\langle\Phi_{\vec z}|\chi_{R/2,\xi}|\Phi_{\vec z}\rangle\right) \\ &
\leq \frac{2 \pi} 3 R^3\rho_\omega\int_{\Bb}d\xi\, 
\langle\Phi_{\vec z}|\chi_{R/2,\xi}|\Phi_{\vec z}\rangle\,.
\end{align}
The last expression can be bounded from above using (\ref{pbe}). 
For the last term in (\ref{nonu}), we use Schwarz's inequality,
together with (\ref{pbe}), to
estimate
\begin{align}
  \int_{\Lambda\setminus\Bb}d\xi\, \langle\Phi_{\vec
    z}|\chi_{R/2,\xi}|\Phi_{\vec z}\rangle^2 &\geq \frac 1{|\Lambda|}
  \left( \int_{\Lambda\setminus\Bb}d\xi \, \langle\Phi_{\vec
      z}|\chi_{R/2,\xi}|\Phi_{\vec z}\rangle \right)^2 \\ \nonumber &
  \geq |\Lambda| \frac{\pi^2}{36} R^6 \left( \rho_{\vec z}^2 - \frac
    {2\pi}3 \rho_{\vec z} \rho \frac{16^3}D \left( 1 + D
      R^3\rho\right)\right) \,.
\end{align}
Here we set again $\rho_{\vec z}=|\vec z|^2/|\Lambda|$.

Putting all these estimates together, we have thus derived the lower
bound
\begin{align}\nonumber
  &\int_{\Lambda\setminus\Bb} d\xi\,
  \Tr_\Ff\big[n_{R/2,\xi}(n_{R/2,\xi}-1)\Omega_b^{\vec z}\big]\\
  \nonumber & \geq |\Lambda| \frac {\pi^2 R^6}{36} \rho_\omega^2 \left
    ( 1 - \frac{1}{D^2} \frac{\pi^2}{9} \left( 1 + D
      R^3\rho\right)\right) + \int_{\Lambda} d\xi \, \tr\,
  \chi_{R/2,\xi}\omega_b\chi_{R/2,\xi}\omega_b \\ \nonumber & \quad +
  |\Lambda| \frac{\pi^2 R^6}{36} \left( 2 \rho_\omega \rho_{\vec z} +
    \rho_{\vec z}^2 - \frac {2\pi}3 \rho_{\vec z} \rho \frac{16^3}D
    \left( 1 + D R^3\rho\right)\right) \\ & \quad - |\Lambda| \frac{2
    \pi^3 R^6}{3 } \rho_\omega \rho \frac{16^3}{18 D} \left( 1 + D
    R^3\rho\right) +2 \int_{\Lambda}d\xi\, \langle\Phi_{\vec
    z}|\chi_{R/2,\xi}\omega_b\chi_{R/2,\xi}|\Phi_{\vec z}\rangle \,.
  \label{waw}
\end{align}
The first integral on the right side of (\ref{waw}) 
can be rewritten as 
\begin{equation}\label{stl}
|\Lambda| \frac{ \pi R^3}{144}  \int_{\Lambda} dx\, |\omega_b(x)|^2
j(d(x,0)/R) \geq |\Lambda| \frac{ \pi^2 R^6}{36} \gamma_b^2 \,,
\end{equation}
where we introduced the notation
\begin{equation}\label{defgb}
\gamma_b = \frac 1{ 4 \pi R^3 } \int_{\Lambda}
   dx\, \omega_b(x) j(d(x,0)/R) \,.
\end{equation}
Eq.~(\ref{stl}) follows by applying Schwarz's inequality to the
integration over $\Lambda$, noting that $\int_{\Lambda} dx\,
j(d(x,0)/R) = 4\pi R^3$.

It remains to integrate the last term in (\ref{waw}). 
We claim that
\begin{equation} \label{cl2}
\int_{\Lambda}d\xi\,\langle\Phi_{\vec
   z}|\chi_{R/2,\xi}\omega_b\chi_{R/2,\xi}|\Phi_{\vec z}\rangle \geq
|\vec z|^2 \frac{ \pi^2 R^6}{36}\left( \gamma_b - R p_c  \rho_\omega\right)
\,.
\end{equation}
To see this, we write
\begin{align}\nonumber
&\frac{144}{\pi R^3}\int_{\Lambda}d\xi\,\langle\Phi_{\vec
   z}|\chi_{R/2,\xi}\omega_b\chi_{R/2,\xi}|\Phi_{\vec z}\rangle- |\vec
z|^2 \int_\Lambda dx\, \omega_b(x) j(d(x,0)/R) \\
\nonumber &= \int_{\Lambda\times\Lambda}dx\,dy\,
\left(\Phi_{\vec z}(x+y)^* - \Phi_{\vec
       z}(y)^*\right) \Phi_{\vec z}(y) \omega_b(x)
j(d(x,0)/R) \\ & \geq - \|\Phi_{\vec z}\|_2
\int_{\Lambda} dx\, \| \Phi_{\vec
   z}(x+\cdot)-\Phi_{\vec z}(\cdot)\|_2 |\omega_b(x)|
j(d(x,0)/R)\,.
\end{align}
We can estimate $|\omega_b(x)|\leq \omega_b(0)=\rho_\omega$. Moreover,
writing the 2-norm as a sum in momentum space, and using the fact that
$\Phi_{\vec z}$ has non-vanishing Fourier coefficients only for $|p|<
p_c$, it is easy to see that $ \| \Phi_{\vec z}(x+\cdot)-\Phi_{\vec
  z}(\cdot)\|_2 \leq \|\Phi_{\vec z}\|_2 p_c d(x,0)$. Since the range
of $j(\,\cdot\,/R)$ is $R$, the integral over $\Lambda$ can be
estimated as $\int_\Lambda dx\, j(d(x,0)/R) d(x,0) \leq R \int_\Lambda
dx\, j(d(x,0)/R) = 4\pi R^4$. This yields (\ref{cl2}).

Collecting all the estimates above, we conclude the following lower
bound on the expectation value of $\dW$:
\begin{align}\nonumber
  \Tr_\Ff\big[ \dW \Upsilon_\pi^{\vec z}\big] & \geq 24
  \frac{\widetilde a}{R^3} \Tr_\Ff\big[ \dN \big( \Upsilon_\pi^{\vec
    z}-\Omega_b^{\vec z}\big)\big]\left( \frac
    1{125}\frac{a'}{\widetilde a} - 6^3 c \frac{R^2}{\eps s^2}\right)
  \\ \nonumber & \quad - \frac {144}\pi \frac {\widetilde a}{R^6}
  \left( b^3 |\Lambda| S(\Upsilon_\pi^{\vec z},\Omega_b^{\vec
      z})\right)^{1/2} \left( 8 + (1+6^3)^2 c \frac{R^2}{\eps
      s^2}\right) \\ \nonumber & \quad - 4\pi \widetilde a |\Lambda|
  \left( 8 \rho_\omega^2 c \frac{R^2}{\eps s^2} + \frac{6^4}\pi
    \frac{\rho_{\vec z} + \rho_\omega}{\eps R s^2} \int_{b/s}^\infty
    dr\, r^6 |m(r)|\right) \\ &\quad + 4\pi a' |\Lambda| \min\left\{
    \Aa_1 \, , \, \Aa_2 \right\} \,. \label{sumw3}
\end{align}
Here we have used the simple bound $a'\leq \widetilde a$, and we have set
\begin{equation}
\Aa_1 = 2 \rho^2 \left(1- \frac {4\pi}3
   \left(\frac R {10}\right)^3\rho_\omega - 8 c \frac{\widetilde a}{a'}
   \frac{6^6 R^2}{\eps s^2} \right)
\end{equation}
and 
\begin{align}\nonumber
  \Aa_2 &= \left( \rho_{\vec z}^2 + 2 \rho_{\vec z} \gamma_b +
    \gamma_b^2\right) + 2 \rho_\omega\rho_{\vec z}\left( 1 - R
    p_c\right) \\ \nonumber &\quad + \rho_\omega^2 \left( 1 -
    \frac{1}{D^2} \frac{\pi^2}{9} \left( 1 + D R^3\rho\right) - 2\pi\,
    3^8 R^3\rho_\omega\right) \\ \nonumber &\quad - \rho_\omega \rho\,
  \frac{4\pi}{3} \frac{16^3}{D} \left(1+D R^3 \rho\right) - \rho^2 \,
  16 c \frac{\widetilde a}{a'} \frac{6^6 R^2}{\eps s^2} \\ &\quad -
  \rho_{\vec z} \rho \, \left( \frac{2^3 3^4}{\pi} \left(16^3
      D\right)^2 R^3 \rho + \frac{2\pi}{3} \frac{16^3}{D} \left( 1 + D
      R^3\rho\right) \right)\,.
\end{align}
We will choose $D=(R^3\rho)^{-1/3}$ in order to minimize the error
terms in $\Aa_2$. Moreover, since $a'/\widetilde a$ contains a factor
$(1-\eps)$ (see (\ref{defap})), it is best to choose $\eps=R/s$. We
note that one can also use the simple bound (\ref{zap}) in order to
estimate $\rho_{\vec z}$ in the error terms.

Since $R_0\ll R\ll s$, the term in round brackets in the first line of
(\ref{sumw3}) is non-negative and, therefore, we will need a lower
bound on $\Tr_\Ff\big[ \dN \big( \Upsilon_\pi^{\vec z}-\Omega_b^{\vec
   z}\big)\big]$. Moreover, we will need an upper bound on the relative
entropy $S(\Upsilon_\pi^{\vec z},\Omega_b^{\vec z})$. Appropriate
bounds will be
derived in the next two subsections.

\subsection{A Bound on the Number of Particles}\label{numbsec}

Our lower bound on the expectation value of $\dW$ in the previous
subsection contains the expression $\Tr_\Ff\big[ \dN \big(
\Upsilon_\pi^{\vec z}-\Omega_b^{\vec z}\big)\big]$, multiplied by a
positive parameter. Hence we need a lower bound on this expression in
order to complete our bound. In fact, we will combine the first term
on the right side of (\ref{sumw3}) with the last term $\half
\Tr_\Ff[\dK\Upsilon^{\vec z}]$ in (\ref{flbF}), which we have not used
so far. I.e., we seek a lower bound on
\begin{equation}
24 \frac{\widetilde a}{R^3} \Tr_\Ff\big[ \dN \big( \Upsilon_\pi^{\vec
    z}-\Omega_b^{\vec z}\big)\big]\left( \frac
    1{125}\frac{a'}{\widetilde a}
    - 6^3 c \frac{R}{s}\right) +  \frac{2\pi \widetilde a
   C}{|\Lambda|} \Tr_{\Ff}\big[ \left(\dN -N\right)^2 \Upsilon^{\vec
   z}\big] \,. \label{num}
\end{equation}
(Here we have used that $\eps=R/s$, as mentioned at the end of the
previous subsection.)  First, note that $\Tr_\Ff
\big[\dN \Omega_{b}^{\vec z}\big] = |\vec z|^2 + \Tr_\Ff\big[ \dN
\Omega_b\big]=|\vec z|^2 + \Tr_\Ff\big[ \dN \Omega_\pi\big] $ and
$\Tr_\Ff \big[\dN \Upsilon_{\pi}^{\vec z}\big] = |\vec z|^2 +
\Tr_\Ff\big[ \dN \Upsilon_\pi\big]$.  Let $\dN^>=\sum_{|p|\geq p_c}
\ad{p}\an{p}$ denote the number operator on $\Ff_>$. Using that
$\Omega_\pi = \Pi\otimes \Gamma^{0}$ and that
$\Upsilon_\pi=\Pi\otimes\Gamma_{\vec z}$, we can thus write
\begin{equation}
\Tr_{\Ff} \big[\dN (\Upsilon_\pi^{\vec z}- \Omega_b^{\vec z})\big] 
= \Tr_{\Ff_>} \big[ \dN^>
(\Gamma_{\vec z}-\Gamma^0)\big] \,.
\end{equation}
For the second term in (\ref{num}), we use 
\begin{align}
  (\dN-N)^2 & \geq \left(|\vec z|^2
    +\Tr_{\Ff_>}\big[\dN^>\Gamma^0\big]- N\right)^2 \\ \nonumber &
  \quad + 2 \left(|\vec z|^2 + \Tr_{\Ff_>}\big[\dN^> \Gamma^0\big] -
    N\right)\left(\dN - |\vec z|^2 - \Tr_{\Ff_>}\big[\dN^>
    \Gamma^0\big]\right)\,,
\end{align} 
and hence
\begin{align}
  \Tr_\Ff\big[(\dN-N)^2 \Upsilon^{\vec z}\big] & \geq \left(|\vec z|^2
    +\Tr_{\Ff_>}\big[\dN^>\Gamma^0\big]- N\right)^2 \\ \nonumber
  &\quad + 2 \left(|\vec z|^2 + \Tr_{\Ff_>}\big[\dN^> \Gamma^0\big] -
    N\right)\Tr_{\Ff_>}\big[\dN^> \left( \Gamma_{\vec
      z}-\Gamma^0\right)\big]\,.
\end{align} 
Thus, we conclude that the expression (\ref{num}) is bounded from below
by 
\begin{align}\label{num2}
  & \frac{2\pi \widetilde a C}{|\Lambda|}\left(|\vec z|^2 +
    \Tr_{\Ff_>}\big[\dN^>\Gamma^0\big]- N\right)^2 + \Tr_{\Ff_>}\big[
  (\dN^>-N^0)\Gamma_{\vec z}\big] \\ & \qquad \times
  \left[\frac{24}{R^3} \left( \frac {a'}{125} - 6^3 c \frac{\widetilde
        a R}{s} \right) + \frac {4\pi \widetilde a C}{|\Lambda|}\left(
      |\vec z|^2 +\Tr_{\Ff_>}\big[\dN^>\Gamma^0\big]- N \right)
  \right] .  \nonumber
\end{align}

We will choose $R$, $s$ and $C$ below in such a way that $R^3\rho \ll
1/C$ and $R\ll s$.  The last term in square brackets is thus positive,
irrespective of the value of $|\vec z|$. Hence we need to derive a
lower bound on $\Tr_{\Ff_>}\big[ \dN^> \left(\Gamma_{\vec
     z}-\Gamma^0\right)\big]$.  To this end, we note that, for any
$\mu\leq 0$,
\begin{equation}\label{varft}
S(\Gamma_{\vec z},\Gamma^0) - \beta \mu\, \Tr_{\Ff_>}\big[ \dN^>
\Gamma_{\vec z}\big] \geq \beta \big(\widetilde f(\mu)-\widetilde
f(0)\big)\,.
\end{equation}
Here, we denoted
\begin{equation}
\widetilde f(\mu) = \frac 1\beta \sum_{|p|\geq p_c} \ln \left( 1-
   e^{-\beta (p^2 - \mu_0-\mu)}\right)\,.
\end{equation}
Eq.~(\ref{varft}) is simply the variational principle for the free
energy $\widetilde f(\mu)$.  Since $\widetilde f$ is completely
monotone (i.e., all derivatives are negative), we can estimate
$\widetilde f(\mu) \geq \widetilde f(0) + \mu \widetilde f'(0) +\half
\mu^2 \widetilde f''(0)$. But $\widetilde f'(0)=-\Tr_{\Ff_>}\big[\dN^>
\Gamma^0\big] $ and hence, optimizing over all (negative) $\mu$,
\begin{equation}
\Tr_{\Ff_>}\big[\dN^>\left(\Gamma_{\vec z}-\Gamma^0\right)\big] \geq 
- \left( S(\Gamma_{\vec z},\Gamma^0) \sum_{|p|\geq p_c}
   \frac 1{\cosh (\beta (p^2-\mu_0) )-1}\right)^{1/2}\,. \label{nnn}
\end{equation}
We can use (\ref{asss}) to estimate the relative entropy as $S(\Gamma_{\vec
   z},\Gamma^{0})\leq 8\pi |\Lambda| \widetilde a\beta\rho^2$.

For the sum over
$p$, we use that $\cosh(x)-1 \geq x^2/2$. In the thermodynamic limit,
we can replace the sum over $p$ by an integral. We thus have to bound
\begin{equation}\label{sni}
\frac 2 {\beta^2} \frac {|\Lambda|}{(2\pi)^3} \int_{|p|\geq p_c} dp\, \frac
1{(p^2 - \mu_0)^2}\,.
\end{equation}
We use two different
bounds. On the one hand,
\begin{equation}
\int_{|p|\geq p_c} dp\, \frac 1{(p^2 - \mu_0)^2}  
\leq  \int_{\R^3} dp\, \frac 1{(p^2-\mu_0)^2} =
\frac{\pi^2}{\sqrt{-\mu_0}} \,.
\end{equation}
On the other hand, since $\mu_0\leq 0$,
\begin{equation}
\int_{|p|\geq p_c} dp\, \frac
1{(p^2 - \mu_0)^2}\leq  \int_{|p|\geq p_c} dp\, \frac 1{(p^2)^2} = \frac{4
   \pi}{p_c} \,.
\end{equation}
In combination, these bounds imply that
\begin{equation}\label{21212}
(\ref{sni}) \leq \frac {|\Lambda|}{\pi^2\beta^2}
\left(  \frac 1{\max\{p_c, 4\pi^{-1}\sqrt{-\mu_0}\}}\right)\,.
\end{equation}

Using (\ref{21212}) in (\ref{nnn}), we obtain the lower bound 
\begin{equation}
\Tr_{\Ff_>}\big[\dN^>\left(\Gamma_{\vec z}-\Gamma^0\right)\big] 
\geq -\const |\Lambda|
\frac{\rho \widetilde a^{1/2}}{\beta^{1/2}} 
\left(p_c +\sqrt{-\mu_0}\right)^{-1/2}
-o(|\Lambda|)\,.
\end{equation}
We apply this bound in (\ref{num2}), noting again that the last term
in square 
brackets is positive. We conclude that 
\begin{equation}\label{nb}
(\ref{num}) \geq  \frac{2\pi \widetilde a C}{|\Lambda|}\left(|\vec z|^2 +
   \Tr_{\Ff_>}\big[\dN^>\Gamma^0\big]- N\right)^2 - Z^{(3)} -
o(|\Lambda|)\,,
\end{equation}
with
\begin{equation} \label{defz3}
Z^{(3)}  =  \const  |\Lambda|
\frac{\rho \widetilde a^{3/2}}{\beta^{1/2}} 
\left(p_c +\sqrt{-\mu_0}\right)^{-1/2}
   \left[\frac{1}{R^3}
      +   C \left(  \rho \left( 1 + 2/ \sqrt{C}\right) 
+ \rho_\omega\right) \right] .
\end{equation}
Here we have used (\ref{zap}) to bound $|\vec z|^2$ from above, as
well as the fact that $\Tr_{\Ff_>}\big[\dN^> \Gamma^0\big] \leq
\Tr_\Ff\big[\dN \Omega_\pi\big] = |\Lambda|\rho_\omega$.

\subsection{Relative Entropy, Effect of Cutoff}\label{relesec}

Next, we are going to give an estimate on the relative entropy of the
two states $\Upsilon_\pi^{\vec z}$ and $\Omega_b^{\vec z}$. This is
needed for the lower bound on the expectation value of $\dW$ obtained
in (\ref{sumw3}). As already noted, the relative entropy is invariant under
unitary transformations, and hence
\begin{equation}
S(\Upsilon^{\vec z}_\pi,\Omega_b^{\vec z})
=S(\Pi\otimes \Gamma_{\vec z},\Omega_b)\,.
\end{equation}
We want to bound this expression by the relative entropy of
$\Pi\otimes \Gamma_{\vec z}$ with respect to $\Omega_\pi=\Pi\otimes
\Gamma^0$, which satisfies
\begin{equation}
S(\Pi\otimes \Gamma_{\vec z},\Omega_\pi) = S(\Gamma_{\vec z},\Gamma^0)
\leq 8\pi \widetilde a \beta |\Lambda|\rho^2
\end{equation}
according to (\ref{asss}). I.e., we want to estimate the effect of the
cutoff $b$ on the relative entropy $S(\Pi\otimes \Gamma_{\vec
  z},\Omega_b)$. Here it will be important that $\Pi$ is not the
vacuum state. The cutoff $b$ corresponds to a mollifying of the
one-particle density matrix in momentum space, and the error in doing
so would not be small enough if this one-particle density matrix is
strictly zero for $|p|<p_c$. This is the reason for replacing the
vacuum state $\Pi_0$ by a more general quasi-free state $\Pi$ in
Subsection~\ref{replsec}.

If $\Omega_\omega$ denotes a general (particle number conserving)
quasi-free state with one-particle density matrix $\omega$, it is easy
to see that $S(\Gamma,\Omega_\omega)$ is convex in $\omega$ for an
arbitrary state $\Gamma$. The one-particle density matrix of
$\Omega_b$ can be written as 
\begin{equation}
\omega_b = \frac 1{|\Lambda|} \sum_q
\widehat \eta_b(q)  \sum_p \half(\omega_\pi(p+q)+\omega_\pi(p-q))
|p\rangle\langle p|\,. 
\end{equation}
Therefore,
\begin{equation}\label{subr}
S(\Pi\otimes \Gamma_{\vec z},\Omega_b) \leq \frac 1{|\Lambda|} \sum_q
\widehat \eta_b(q) S(\Pi\otimes\Gamma_{\vec z}, \Omega_{q})\,,
\end{equation}
where $\Omega_{q}$ is the quasi-free state corresponding to the
one-particle density matrix with eigenvalues
$\half(\omega_\pi(p+q)+\omega_\pi(p-q))$. (This is the same estimate
as in \cite[Sect.~5.2]{RSjellium}.)  Moreover, as in
\cite[Eq.~(5.14)]{RSjellium}, simple convexity arguments yield
\begin{align}\label{expe}
S(\Pi\otimes\Gamma_{\vec z}, &\Omega_{q}) \leq  \left(1+t^{-1}\right)
S(\Pi\otimes\Gamma_{\vec z},\Omega_\pi) \\ \nonumber &+ \sum_p
\big(h_q(p)-h_0(p)\big)\left( \frac{1}{e^{h_0(p)+t(h_0(p)-h_q(p))}-1}-
    \frac{1}{e^{h_q(p)}-1}\right)
\end{align}
for any $t>0$. Here
\begin{equation}\label{hqp}
h_q(p)=\ln \frac{2+\omega_\pi(p+q)+\omega_\pi(p-q)}
{\omega_\pi(p+q)+\omega_\pi(p-q)}\,.
\end{equation}
Note that $h_0(p)=\ell(p)$, defined in (\ref{defell1}). To estimate the
expression (\ref{expe}) from above, we need the following lemma.

\begin{lem}\label{lemhq}
  Let $\ell: \R^3\mapsto \R_+$, and let $L_\pm = \pm \sup_{p} \sup_{q,
    \|q\|=1} \pm (q\nabla)^2\ell(p)$ denote the supremum (infimum) of
  the largest (lowest) eigenvalue of the Hessian of $\ell$. Let
  $\omega_\pi(p)=[ e^{\ell (p)} -1]^{-1}$, and let $h_q(p)$ be given
  as in (\ref{hqp}). Then
\begin{equation}\label{lemup}
h_q(p) - h_0(p) \leq  L_+ q^2\,,
\end{equation}
and 
\begin{align}\nonumber
  h_q(p) - h_0(p) &\geq q^2 L_- + q^2 \min\{L_-,0\} - 4 q^2 \sup_p[
  |\nabla \ell(p)|^2 \omega_\pi(p)] \\ & \quad - 2 q^2 (|q|+|p|)^2
  \sup_p [|\nabla \ell(p)|^2/p^2] \,.\label{lemdown}
\end{align}
\end{lem}

\begin{proof} 
By convexity of $x\mapsto \ln(1+1/x)$,
\begin{equation}
h_q(p) \leq \half ( \ell(p+q) + \ell(p-q) ) \leq \ell(p) + L_+ q^2\,,
\end{equation}
proving the (\ref{lemup}). To obtain the lower bound in
(\ref{lemdown}), we proceed similarly to \cite[Lemma~5.2]{RSjellium}.
We can write
\begin{equation}\label{useb}
h_q(p)-h_0(p) = \int_0^1 d\lambda\, (1-\lambda) \frac
{\partial^2}{\partial\lambda^2} h_{\lambda q}(p)\,.
\end{equation}
Denoting $p_{\pm}=p\pm \lambda q$ and $\omega_\pm=\omega_\pi(p_\pm)$, we
can write the second derivative of $h_{\lambda q}(p)$ as
\begin{align} \nonumber
  \frac {\partial^2}{\partial\lambda^2} h_{\lambda q}(p)=&\left[\frac
    1{(\omega_++\omega_-)^2}-\frac{1}
    {\left(2+\omega_++\omega_-\right)^2}\right]
  \left( \frac\partial{\partial\lambda} (\omega_++\omega_-)\right)^2 \\
  &- \frac 2{(\omega_++\omega_-)(2+\omega_++\omega_-)}
  \frac{\partial^2}{\partial\lambda^2}(\omega_++\omega_-)\,.
\end{align}
The first term is positive and can thus be neglected for a lower
bound. For the second term, we use
\begin{equation}
\frac
{\partial^2}{\partial\lambda^2} \omega_+ =
\omega_+(1+\omega_+)\left[ (1+2\omega_+)\left( q\nabla \ell_+\right)^2 -
     (q\nabla)^2\ell_+\right]\,,
\end{equation}
where we denoted $\ell_+=\ell(p+\lambda q)$.  The last term in the
square brackets is bounded above by $ - q^2 L_-$. Moreover,
\begin{align}\nonumber
  (1+&2\omega_+)\left( q\nabla \ell_+\right)^2 \\ & \leq q^2
  (|p|+|q|)^2 \sup_p \left[ |\nabla\ell(p)|^2/p^2\right] + 2 q^2
  \sup_p \left[ \omega_\pi(p)|\nabla\ell(p)|^2\right]\,.
\end{align}
The same bounds hold with $+$ replace by $-$. Using in addition that
\begin{equation}
\frac 12  \leq \frac{\omega_+(1+\omega_+) +
   \omega_-(1+\omega_-)}{(\omega_++\omega_-)(2+\omega_++\omega_-)} \leq
1\,,
\end{equation}
we arrive at (\ref{lemdown}).
\end{proof}

Let $g:\R^3\mapsto [0,1]$ be a smooth radial function supported in  the
ball of radius $1$. We assume that $g(p)\geq \half $ for $|p|\leq
\half$. We then choose 
\begin{equation}\label{defell}
\ell(p) = \beta ( p^2 - \mu_0)  + \beta p_c^2 g (p/p_c)\,.
\end{equation}
Since $\ell(p)=\ln (1+1/\pi_p)$ for $|p|<p_c$ by definition, this
corresponds to the choice $\pi_p= (\exp(\beta ( p^2 - \mu_0) + \beta
p_c^2 g (p/p_c))-1)^{-1}$. Note that, in particular, $\pi_p\leq \const
/(\beta p_c^2)$ and hence $P \leq \const M/(\beta p_c^2)\sim p_c
|\Lambda|/\beta$. This bound is important for estimating the error term
$Z^{(2)}$ in (\ref{defz2}).

For the $\ell$ given in (\ref{defell}), both $\beta^{-1}L_+$ and
$\beta^{-1} L_-$ are bounded independent of all parameters. Moreover
$|\nabla \ell(p)|\leq \const \beta |p|$. Using that $\omega_\pi(p)\leq
\ell(p)^{-1} \leq (\beta p^2)^{-1}$, the bounds in Lemma~\ref{lemhq}
imply that
\begin{equation}\label{lbo}
- D \beta q^2 \left ( 1 +   \beta
   (|p|+|q|)^2 \right) \leq h_q(p) - h_0(p)  \leq D \beta q^2 
\end{equation}
for some constant $D>0$.

Using the fact that $\sinh(x)/x \leq \cosh(x)$ for all $x\in \R$, we
can estimate
\begin{align} \label{lll}
  &\big(h_q(p)-h_0(p)\big)\left(
    \frac{1}{e^{h_0(p)+t(h_0(p)-h_q(p))}-1}-
    \frac{1}{e^{h_q(p)}-1}\right) \\ &\leq \half (1+t)
  \big(h_q(p)-h_0(p)\big)^2 \frac{ e^{-h_q(p)} +
    e^{-h_0(p)+t(h_q(p)-h_0(p))}}{\big(1- e^{-
      h_0(p)+t(h_q(p)-h_0(p))}\big)\big(1- e^{-h_q(p)}\big)}\,.
  \nonumber
\end{align}
The estimate (\ref{lbo}) implies the bound
\begin{equation}\label{lll2}
\big(h_q(p)-h_0(p)\big)^2 \leq D^2 (\beta q^2)^2 \left ( 1 +
   \beta
   (|p|+|q|)^2 \right)^2\,.
\end{equation}
Moreover, using the upper bound in
(\ref{lbo}) to estimate $h_q(p)-h_0(p)$ from above, we obtain
\begin{align}\nonumber
&\frac{ e^{-h_q(p)} +
     e^{-h_0(p)+t\beta D q^2}}{\big(1- e^{-
       h_0(p)+t \beta D q^2}\big)\big(1- e^{-h_q(p)}\big)} \\ & \qquad
   =
   \omega^t(p) +   \half\big(\omega_\pi(p+q)+\omega_\pi(p-q)\big)
   \left( 1 + 2 \omega^t(p)\right) 
\end{align}
as an upper bound to the last fraction in (\ref{lll}). Here, we
denoted $\omega^t(p) = [ e^{h_0(p)- D\beta t q^2} -1]^{-1}$, assuming
$t$ to be small enough such that $h_0(p)- D\beta t q^2> 0$ for all
$p$. Recall that $h_0(p)=\ell(p)$ is given in (\ref{defell}).

In the thermodynamic limit, the sum over $p$ in (\ref{expe}) converges
to the corresponding integral and, hence, we are left with bounding
\begin{equation}\label{integg}
\int_{\R^3} dp\, \left ( 1 +   \beta
   (|p|+|q|)^2 \right)^2 \left( \omega^t(p) +   
\half\big(\omega_\pi(p+q)+\omega_\pi(p-q)\big)
   \left( 1 + 2 \omega^t(p)\right) \right)
\end{equation}
from above. We can replace
$\omega_\pi(p-q)$ by
$\omega_\pi(p+q)$ in (\ref{integg}) without changing the value of the
integral. Using Schwarz's inequality, the fact that $\omega_\pi(p) \leq
\omega^t(p)$, and changing variables
$p\to p-q$, we see that (\ref{integg}) is bounded from above by
\begin{equation}\label{inter2}
2 \int_{\R^3} dp\, \left ( 1 +
   \beta
   (|p|+ 2 |q|)^2 \right)^2  \omega^t(p)\big(1 + 2\omega^t(p)\big)\,.
\end{equation}

It remains to bound $\omega^t(p)$. For this purpose, we need a bound
on $\ell(p) - D\beta t q^2$ for an appropriate choice of the parameter
$t$.  We will choose $t=\min\{1, (b^2 q^2)^{-1}\}$. We then have $t
q^2 \leq b^{-2}$, and it is easy to see that
\begin{equation}
\ell(p) - D\beta t q^2 \geq \beta \left[ \half p^2 - \mu_0 +
    p_c^2 \left( \frac 18 - \frac {D}{b^2 p_c^2}\right)
\right]
\end{equation}
in this case. Since $\ell(p)\geq \beta(p^2 -\mu_0)$, this can be seen
immediately in the case $|p| \geq \half p_c$. For $|p|\leq p_c/2$ even
a slightly better bound holds, this time using the fact that $g(p)
\geq 1/2$ for $|p|\leq 1/2$.

We will choose $b$ and $p_c$ below in such that a way that $b p_c \gg
1$. Denoting by $\tau$ the (positive) number
\begin{equation}
\tau = - \beta \mu_0 + \beta  p_c^2 \left( \frac 18 - \frac
   {D}{b^2 p_c^2}\right)\,,
\end{equation}
we thus have the bound
\begin{equation}
\omega^t(p) \leq \left[ e^\tau e^{\half\beta p^2} -1\right]^{-1} \leq
   e^{-\tau} e^{-\half\beta p^2} \left( 1 + \frac 1{\tau + \half
        \beta p^2}\right) \,.
\end{equation}
The last bound follows from the elementary inequality $(e^x-1)^{-1}
\leq e^{-x}(1+1/x)$ for all $x >0$.
We insert this bound for $\omega^t$ into (\ref{inter2}). Simple
estimates then yield 
\begin{equation}\label{lll3}
(\ref{inter2})\leq \const \frac {e^{-\tau}}{\beta^{3/2}} \left( 1+
   \tau^{-1/2}\right)\left( 1+ (\beta q^2)^2\right)\,. 
\end{equation}

Combining (\ref{expe}), (\ref{lll}), (\ref{lll2}) and (\ref{lll3}),
and using that $t^{-1} \leq 1 + b^2 q^2$, we have thus shown that 
\begin{align}\nonumber
S(\Pi\otimes \Gamma_{\vec z},\Omega_q) \leq &\left( 2 + b^2 q^2 \right)
S(\Pi\otimes \Gamma_{\vec z},\Omega_\pi) \\ & + \const |\Lambda|
\beta^{1/2} q^4\tau^{-1/2}
\left(1+(\beta q^2)^2\right) +
o(|\Lambda|).
\end{align}
After inserting this estimate in (\ref{subr}) and summing over $q$, this
yields the bound 
\begin{equation}
S(\Upsilon_\pi^{\vec z},\Omega_b^{\vec z})\leq \const |\Lambda| \left(
   \widetilde a\beta\rho^2  + 
\frac{\beta^{1/2}\tau^{-1/2}}{b^4} \right)
+ o(|\Lambda|). \label{ssb}
\end{equation}
Here, we have used the assumed smoothness of $\eta$ to estimate
$\sum_q \widehat \eta_b (q) |q|^n \leq \const | \Lambda| b^{-n}$ for
integers $n\leq 8$. We have also used the fact that we will choose
$b^{2} \gg \beta$ and hence, in particular, $\beta b^{-2} \leq \const$
Moreover, the assumption (\ref{asss}) has been used to bound
$S(\Pi\otimes\Gamma_{\vec z},\Omega_\pi) = S(\Gamma_{\vec
  z},\Gamma^0)$.

We have thus shown that the effect of the cutoff $b$ on the relative
entropy can be estimated by $|\Lambda|(\beta/\tau)^{1/2} b^{-4}$. We
note that the exponent $-4$ of $b$ is important, since the relative
entropy has to be multiplied by $b^3$ in the estimate (\ref{sumw3}).

\subsection{Final Lower Bound on 
${\mathord{\hbox{\boldmath $F_{\vec z}(\beta)$}}}$}

We have now obtained all the necessary estimates to complete our lower
bound on $F_{\vec z}(\beta)$. For this purpose, we put all the bounds
from Subsections~\ref{holesec}, \ref{intesec}, \ref{numbsec}
and~\ref{relesec} together. In fact, from (\ref{flbF}), (\ref{sumw3}),
(\ref{nb}) and (\ref{ssb}), we have the following lower bound on
$F_{\vec z}(\beta)$:
\begin{align} \nonumber
  F_{\vec z}(\beta) \geq & -\frac 1\beta \ln \Tr_{\Ff_>} \exp\big(
  -\beta \dT_{\rm s}^c(\vec z)\big) - Z^{(2)} - Z^{(3)} - Z^{(4)} \\ 
  \nonumber &- (\kappa-\kappa')\sum_{|p|<p_c} p^2\pi_p - o(|\Lambda|)
  \\ \nonumber & + \frac{2\pi \widetilde a C}{|\Lambda|}\left(|\vec
    z|^2 + \Tr_{\Ff_>}\big[\dN^>\Gamma^0\big]- N\right)^2 \\ &+ 4\pi
  \widetilde a |\Lambda| \min\left\{ 2 \rho^2\, , \, \rho_{\vec z}^2 +
    2 \rho_{\vec z}(\gamma_b + \rho_\omega) + \rho_\omega^2
    +\gamma_b^2\right\}\,,\label{bfa}
\end{align}
where we denoted 
\begin{align}\nonumber
  Z^{(4)} & = \const \widetilde a |\Lambda| \Biggl[\rho^2 \left(
    \kappa + \frac Rs + Rp_c + (R^3\rho)^{1/3}+
    \left(\frac{R_0}{R}\right)^3 \right) \\ & \qquad \label{defz4} +
  \frac{\rho}{R^2 s} \int_{b/s}^{\infty} dr\, r^6 |m(r)| + \frac
  1{R^6}\left( b^3 \widetilde a \beta \rho^2 + \frac{\beta^{1/2}
      \tau^{-1/2}}{b}\right)^{1/2} \Biggl]\,.
\end{align}
Here, we have used the definition (\ref{defap}) of $a'$, (\ref{zap})
to bound $\rho_{\vec z}$ in the error terms, together with
$\gamma_b\leq \rho_\omega$ and $\lim_{|\Lambda|\to \infty}\rho_\omega
\leq \rho$. This last estimate follows from $\ell(p)\geq
\beta(p^2-\mu_0)$ and (\ref{eq16}). The error terms $Z^{(2)}$ and
$Z^{(3)}$ are defined in (\ref{defz2}) and (\ref{defz3}),
respectively.

Using (\ref{kappapr}), the term in the second line of (\ref{bfa}) can
be estimated by $(\kappa-\kappa')\sum_{|p|<p_c} p^2\pi_p \leq
(\kappa-\kappa') p_c^2 P \leq \const (\widetilde a/R)^3 p_c^3
|\Lambda| /\beta$. Here, we have also used the bound on
$P=\sum_{|p|<p_c}\pi_p$ derived after Eq.~(\ref{defell}).

The two terms in the third and forth line of (\ref{bfa}) can be
bounded from below independently of $\vec z$, simply using Schwarz's
inequality. More precisely, introducing $\rho^0 \equiv |\Lambda|^{-1}
\Tr_{\Ff_>}\big[\dN^> \Gamma^0\big] = \rho_\omega - P/|\Lambda|$, we
obtain
\begin{align}\nonumber
  & \frac{2\pi \widetilde a C}{ |\Lambda|}\left(|\vec z|^2 +
    \Tr_{\Ff_>}\big[\dN^>\Gamma^0\big]- N\right)^2 + 4\pi \widetilde a
  |\Lambda| \left( \rho_{\vec z}^2 + 2 \rho_{\vec z}(\gamma_b +
    \rho_\omega) + \rho_\omega^2 +\gamma_b^2\right) \\ & \geq \frac
  {4\pi \widetilde a |\Lambda| }{1+ 2/C} \left( (\rho-\rho^0)^2 + 2
    (\rho-\rho^0)(\rho_\omega+\gamma_b) +\rho_\omega^2 + \gamma_b^2 -
    \frac 2C (\rho_\omega+\gamma_b)^2\right)\,. \label{finas}
\end{align}
We note that 
\begin{equation}
\rho^0 = \frac 1{(2\pi)^3} \int_{|p|\geq p_c} dp\, \frac
1{e^{\beta(p^2-\mu_0)}-1} + o(1)
\end{equation}
in the thermodynamic limit. Hence, from (\ref{eq16}),
\begin{align}\nonumber
\rho^0 & = \min\left\{\rho\, , \, \rho_c(T)\right\} -\frac 1{(2\pi)^3} 
\int_{|p|\leq p_c} dp\, \frac
1{e^{\beta(p^2-\mu_0)}-1}  +o(1) \\ & \geq  \min\left\{\rho\, , \,
   \rho_c(T)\right\} - \frac 1{2\pi^2} \frac {p_c}{\beta} +o(1)\,.
\end{align}
This estimate is obtained by using $e^{\beta(p^2-\mu_0)}-1 \geq \beta
p^2$ in the denominator of the integrand. Moreover, $\rho^0\leq
\rho_\omega \leq \min\{\rho,\rho_c(T)\} + o(1)$.

It remains to give a
lower bound on $\gamma_b$. According to (\ref{defgb}) and (\ref{defgd}),
\begin{equation}\label{defgb2}
\gamma_b = \frac 1{ 4 \pi R^3 } \int_{\Lambda}
   dx\, \omega_\pi(x) \eta(d(x,0)/b) j(d(x,0)/R) \,.
\end{equation} 
We note that, since $\omega_\pi(x)$ is real, 
\begin{equation}
\omega_\pi(x) -\rho_\omega = \frac 1{|\Lambda|} \sum_p \frac
1{e^{\ell(p)}-1} \left( \cos(px) -1\right) \geq - \frac{ d(x,0)^2 } {
   2 |\Lambda|} \sum_p \frac {p^2}{e^{\ell(p)}-1}\,.
\end{equation}
We can estimate $d(x,0)\leq R$ in the integrand in
(\ref{defgb2}). Since $\ell(p)\geq \beta |p|^2$, $|\eta|\leq 1$ and
$\int_\Lambda dx\, j(d(x,0)/R)=4\pi R^3$, the contribution of the last
term to (\ref{defgb2}) is bounded by 
\begin{equation}
\frac{R^2}{2} \frac 1{(2\pi)^3 \beta^{5/2}} \int_{\R^3} dp\,
\frac{p^2}{e^{p^2} -1} + o(1)
\end{equation}
in the thermodynamic limit.  Moreover, we can bound $\eta$ from below
as $\eta(t) \geq 1 -\const t^2$, and hence
\begin{equation}
\rho_\omega \geq \gamma_b \geq \rho_\omega  
\left (1-\const \frac{R^2}{b^2}\right) -
\const R^2 \beta^{-5/2} - o(1)\,.
\end{equation}

Using the bounds on $\rho^0$ and $\gamma_b$ just derived, we have
\begin{align}\label{rhss}
  (\ref{finas}) & \geq 4\pi \widetilde a |\Lambda| \left( 2 \rho^2 -
    \left[ \rho-\rho_c(T)\right]_+^2\right) \\ \nonumber & \quad -
  \const \widetilde a |\Lambda|\left( \rho^2 \left( \frac 1{C}
      +\frac{R^2}{b^2}\right) + \rho\left(\frac{p_c}\beta +
      \frac{R^2}{\beta^{5/2}} \right)\right) - o(|\Lambda|)\,.
\end{align}
In particular, the terms in the third and forth line of (\ref{bfa})
are bounded from below by the right side of (\ref{rhss}).

\subsection{The \lq\lq Free\rq\rq\ Free Energy}

We now insert the lower bound on $F_{\vec z}(\beta)$ derived in the
previous subsection into (\ref{eq234}). We note that the only $\vec
z$-dependence left is in the first term in (\ref{bfa}), which is the
\lq\lq free\rq\rq\ part of the free energy. Taking also the constant
$\mu_0 N$ in (\ref{eq234}) into account, we are thus left with
evaluating
\begin{align}\nonumber
  &\mu_0 N - \frac 1\beta \ln \int_{\C^M} d^{M}\!z \,
  \Tr_{\Ff_>}\exp\big( -\beta \dT_{\rm s}^c(\vec z)\big) \\ & = \mu_0
  N + \frac 1\beta \left( \sum_{|p|<p_c} \ln \big(\beta \eps(p)\big) +
    \sum_{|p|\geq p_c} \ln \left( 1- e^{-\beta \eps(p)}\right)
  \right)\,.\label{fff}
\end{align}
Using $x\geq (1-e^{-x})$ for non-negative $x$, (\ref{fff}) becomes, in
the thermodynamic limit,
\begin{equation}
(\ref{fff})\geq  N \mu_0  + \frac {|\Lambda|}{\beta (2\pi)^{3}}
\int_{\R^3}dp\, \ln \left( 1-e^{-\beta \eps(p)}\right) 
- o(|\Lambda|)\,. \label{ffe}
\end{equation}
Recall that $\eps(p)$ is defined in (\ref{ltx3t}). It is given by
$\eps(p)= (1-\kappa+\kappa') p^2 - \mu_0$ for $|p|\leq 1/s$, and
satisfies the bound $\eps(p) \geq \kappa' p^2$ for $|p|\geq 1/s$.
Hence
\begin{align}\nonumber
  \int_{\R^3} dp\, \ln\left( 1 - e^{-\beta \eps(p)}\right) &\geq
  (1-\kappa+\kappa')^{-3/2} \int_{\R^3} dp\, \ln\left( 1 - e^{-\beta
      (p^2- \mu_0)}\right) \\ &\quad + \frac 1{(\kappa'\beta)^{3/2}}
  \int_{|p|^2\geq \kappa'\beta/s^2} dp\, \ln\left( 1 - e^{- p^2
    }\right)\,. \label{skb}
\end{align}
We will choose $s^2 \ll \kappa' \beta$ below, in which case the last
integral is exponentially small in the parameter $s^2/(\kappa'
\beta)$.  Inserting the definition (\ref{fnodeb}) of
$f_0(\beta,\rho)$, we have thus shown that
\begin{align}\label{2154}
  (\ref{fff}) & \geq |\Lambda| (1-\kappa+\kappa')^{-3/2}
  f_0(\beta,\rho) \\ \nonumber & \quad + \frac {|\Lambda|}{\beta^{5/2}
    \kappa'^{3/2} (2\pi)^{3}} \int_{|p|^2\geq \beta\kappa'/s^2} dp\,
  \ln \left(1-e^{-p^2}\right) - o(|\Lambda|)\,.
\end{align}
Here, we have also used that $\mu_0\leq 0$.

\subsection{Choice of Parameters}

We have now essentially finished our lower bound on $f(\beta,\rho)$.
It remains to collect all the error terms, and choose the various
parameters in an appropriate way. All the error terms we have to take
into account are given in (\ref{eqlem1}), (\ref{eq234}), (\ref{bfa}),
(\ref{rhss}) and (\ref{2154}).

We will choose the various parameters in our estimates as follows:
\begin{align}\nonumber
R &= \rho^{-1/3} \big( a \rho^2 \beta^{5/2} \big)^{3/403}\quad , \quad
b = \beta^{1/2} \big( a \rho^2 \beta^{5/2} \big)^{-121/403} \,, \\ 
s &= \big(\beta\rho^{-1/3}\big)^{1/3} \big( a \rho^2 \beta^{5/2}
\big)^{1/403} \,. 
\end{align}
Moreover,
\begin{equation}
\varphi = a \big( a \rho^2 \beta^{5/2} \big)^{-A}\quad , \quad 
C = \big( a \rho^2 \beta^{5/2} \big)^{-B}
\end{equation}
for $4/403 \leq A \leq 79/403$ and $2/403\leq B \leq
161/403$. Depending on $\mu_0$, we choose
\begin{equation}
p_c = \left\{ \begin{array}{ll}
\beta^{-1/2} \big( a \rho^2 \beta^{5/2} \big)^{81/403} & {\rm if\ }
\beta |\mu_0| \leq \big( a \rho^2 \beta^{5/2} \big)^{162/403} \\ 0 &
{\rm otherwise} \,. \end{array}\right.
\end{equation}
Finally, we choose $\kappa = s^2 \beta^{-1} \big( a \rho^2 \beta^{5/2}
\big)^{-\delta}$ for some $\delta>0$. Our estimates then imply that
\begin{equation}\label{finnu}
f(\beta,\rho) \geq f_0(\beta,\rho) + 4\pi a \left( 2 \rho^2 -
   [\rho-\rho_c(\beta)]_+^2\right) \big(1 - o(1)\big) \,,
\end{equation}
with 
\begin{equation}
o(1) \leq C_\delta \big( (\beta\rho^{2/3})^{-1} \big)  
\big( a \rho^2 \beta^{5/2} \big)^{2/403-\delta}
\end{equation}
for some function $C_\delta$, depending on $\delta$, that is uniformly
bounded on bounded intervals.

The choice of the parameters $p_c$, $b$, $s$, $R$ and $\kappa$ is
determined by minimizing the sum of all the error terms. The main
terms to consider are, in fact, the terms $Mp_c^2 \sim |\Lambda|
p_c^5$ in $Z^{(1)}$ in (\ref{defz1}), and $|\Lambda|\widetilde
a\rho^2(\kappa + R/s)$ as well as $|\Lambda| \widetilde a R^{-6}( b^3
\widetilde a \beta \rho^2 + (p_c^2-\mu_0)^{-1/2}b^{-1})^{1/2}$ in
$Z^{(4)}$ in (\ref{defz4}). Moreover, we have to take the restriction
$s^2\ll \kappa\beta$ in (\ref{skb}) into account. This leads to the
choice of parameters above.

All other error terms are of lower order for small
$a\rho^2\beta^{5/2}$. This is true, in particular, for all the terms
containing $\varphi$ and $C$, which explains the freedom in their
choice above.

\subsection{Uniformity in the Temperature}

Our final result, Eq.~(\ref{finnu}), does not have the desired
uniformity in the temperature. It is only useful in case the
dimensionless parameter $a\rho^2\beta^{5/2}$ is small. In particular,
one can not take the zero-temperature limit $\beta\rho^{2/3}\to
\infty$. The reason for this restriction is that our argument was
essentially perturbative, using that the correction term we want to
prove is small compared to the main term, i.e., $a\rho^2 \ll
f_0(\beta,\rho)$. Below the critical temperature, $f_0(\beta,\rho)=
\const \beta^{-5/2}$, hence the assumption is only satisfied if
$a\rho^2\beta^{5/2}\ll 1$.

If the temperature is smaller, we can use a different argument for a
lower bound on $f$, which uses in an essential way the result in
\cite{LY1998}. There, a lower bound in the zero temperature case was
derived.

To obtain the desired bound for very low temperature, it is possible
to skip steps 1--5 entirely, and start immediately with the Dyson
lemma, Lemma~\ref{dlem}, applied to the original potential $v$. Using
this lemma, we have that 
\begin{align}
  H_N & \geq \sum_{j=1}^N \biggl[ -\nabla_j
   (1-(1-\kappa)\chi(p_j)^2) \nabla_j
    \\ & \qquad +(1-\eps)(1-\kappa)  a U_R(d(x_j,x_{\rm
      NN}^{J_j}(x_j))) - \frac a\eps
   \sum_{i\in J_j} w_R(x_j-x_i)\biggl] \,. \nonumber
\end{align}
Since $d(x_i,x_k)\geq R/5$ for $i,k\in J_j$, we can estimate (using
(\ref{wg}))
\begin{equation}
  \sum_{j=1}^N  \frac a\eps  \sum_{i\in J_j} w_R(x_j-x_i) \leq \const
  \frac {a N} {\eps R s^2}\,.
\end{equation} 
Moreover, the calculation in \cite{LY1998} shows that, for the choice
$\kappa = (a^3\rho)^{1/17}$ and $R= a (a^3\rho)^{-5/17}$,
\begin{align}\nonumber
   &\sum_{j=1}^N -\frac \kappa 2 \Delta_j +(1-\eps)(1-\kappa) a
   U_R(d(x_j,x_{\rm NN}^{J_j}(x_j))) \\ &\quad \geq 4\pi a N\rho \left(
     1 - \eps-\const (a^3\rho)^{1/17} \right)\,.
\end{align}
(Strictly speaking, this result was derived in \cite{LY1998} for
Neumann boundary conditions, and with $J_j = \{1,\dots,N\}$
independent of all the particle coordinates. It is easy to see,
however, that the same result applies to our Hamiltonian, being
defined with periodic boundary conditions, and having a slightly
smaller interaction.)

For this choice of $\kappa$ and $R$, we thus have
\begin{equation}
H_N \geq \sum_{j=1}^N l(\sqrt{-\Delta_j}) + 4\pi a N \rho\left(
     1 - \eps-\const (a^3\rho)^{1/17} 
- \const \frac 1 {\eps R s^2 \rho}\right)\,,
\end{equation}
with $l(|p|) = p^2 (1 - \kappa/2 - (1-\kappa)\chi(p)^2)$. To obtain a
lower bound on the free energy for this Hamiltonian, we can go to
grand-canonical ensemble, introducing a chemical potential in the
usual way. Taking the thermodynamic limit, this yields
\begin{align}\nonumber
  f(\beta,\rho) &\geq \sup_{\mu\leq 0} \left\{ \mu \rho+ \frac
    1{(2\pi)^3\beta} \int_{\R^3} dp\, \ln\left( 1 - e^{-\beta (l(p)-
        \mu)}\right) \right\} \\ & \quad + 4\pi a \rho^2 \left( 1 -
    \eps -\const (a^3\rho)^{1/17} \right) - \const \frac {\rho} {\eps
    s^2}(a^3\rho)^{5/17}\,.
\end{align}
Recall that $\chi(p)= \nu(sp)$, where $\nu$ is a function with $0\leq
\nu(p)\leq 1$ that is
supported outside the ball of radius $1$. This
implies that $l(p)=(1-\half\kappa) p^2$ for $|p|\leq 1/s$, and
$l(p) \geq \half \kappa p^2$ for $|p|\geq 1/s$. Hence
\begin{align}\nonumber
\int_{\R^3}
dp\, \ln\left( 1 - e^{-\beta (l(p)- \mu)}\right) &\geq (1-\half
\kappa)^{-3/2} \int_{\R^3}
dp\, \ln\left( 1 - e^{-\beta (p^2- \mu)}\right) \\ &\quad + \frac
1{(\kappa\beta)^{3/2}} \int_{|p|^2\geq \kappa\beta/s^2}
dp\, \ln\left( 1 - e^{-\half p^2 }\right)\,.
\end{align}
The last expression is exponentially small in the (small) parameter
$s^2/(\kappa\beta)$. We choose, for some $\delta>0$, 
\begin{equation}
s^2 = \beta (a^3\rho)^{1/17+\delta} \quad , 
\quad \eps^2 = \frac{ (a^3\rho)^{3/85}}{(a\rho^2\beta^{5/2})^{2/5}}\,,
\end{equation}
and obtain
\begin{equation}\label{fint}
f(\beta,\rho) \geq f_0(\beta,\rho) + 4\pi a \rho^2 \left( 1 - o(1)\right)\,,
\end{equation}
with
\begin{equation}
o(1) = \const \left( (a^3\rho)^{1/17} 
\left(1+\frac{1}{ a\rho^2\beta^{5/2}}\right)
+ \frac{ (a^3\rho)^{3/170-2\delta}}{(a\rho^2\beta^{5/2})^{1/5}} \right)\,. 
\end{equation}
Compared with our desired lower bound, we also have to take into
account the missing term $a(\rho^2 - [\rho-\rho_c(\beta)]_+^2)$, which
can be bounded as 
\begin{equation}
a(\rho^2 - [\rho-\rho_c(\beta)]_+^2)\leq \const  a\rho\beta^{-3/2}\,.
\end{equation}

In combination, the estimates (\ref{finnu}) and (\ref{fint}) provide
the desired uniform lower bound on the free energy. Depending on the
value of $a\rho^2\beta^{5/2}$, one can apply either one of them. To
minimize the error, one has to apply (\ref{finnu}) for
$a\rho^2\beta^{5/2} \leq (a^3\rho)^{403/6885}$, and (\ref{fint})
otherwise. This yields our main result, Theorem~\ref{T1}, for $\alpha=
2/2295 - \delta$.

\appendix
\numberwithin{equation}{section}

\section{Appendix: Proof of Lemma~\ref{dlem}}

For simplicity, we drop the $\widetilde{\phantom{v}}~$ on $v$ and $a$
in our notation in this appendix.

We start by dividing up $\Lambda$ into Voronoi cells
\begin{equation}
\Bb_j = \{ x\in \Lambda\, : \, d(x,y_j) \leq d(x,y_k) \ \forall k\neq
j\}\,.
\end{equation}
For a given $\psi \in H^1(\Lambda)$, let $\xi$ be the function with
Fourier coefficients $\widehat \xi(p) = \chi(p) \widehat \psi(p)$. We
thus have to show that
\begin{align}\nonumber
  \int_{\Bb_j}dx\, \left[ |\nabla \xi(x)|^2 + \half v(d(x,y_j))
    |\psi(x)|^2\right] & \geq (1-\eps) a \int_{\Bb_j}dx\, U(d(x,y_j))
  |\psi(x)|^2 \\ & \quad - \frac a \eps \int_\Lambda dx\, w_R(x-y_j)
  |\psi(x)|^2\,.  \label{toshow}
\end{align}
The statement of the lemma then follows immediately by summing over
$j$ and using the positivity of $v$.

We will actually show that (\ref{toshow}) holds even when the
integration region $\Bb_j$ on the left side of the inequality is
replaced by the smaller set $\Bb_R \equiv \Bb_j \cap \{ x\in\Lambda :
d(x,y_j)\leq R\}$. Note that the first integral on the right side of
(\ref{toshow}) is also over this region, since the range of $U$ is
supposed to be less than $R$.

As in Subsection~\ref{defas}, let $\phi_v$ denote the solution to the
zero-energy scattering equation
\begin{equation}\label{scatteq}
-\Delta \phi_v(x) + \half v(|x|) \phi_v(x) = 0
\end{equation}
subject to the boundary condition $\lim_{|x|\to \infty} \phi_v(x)=1$.
Let $\nu$ be a complex-valued function on the unit sphere $\Ss^2$,
with $\int_{\Ss^2} |\nu|^2 = 1$.  We use the same symbol for the
function on $\R^3$ taking values $\nu(x/|x|)$.  For $\psi$ and $\xi$
as above, consider the expression
\begin{align}\nonumber
  A & \equiv \int_{\Bb_R}dx\, \nu(x-y_j) \nabla\xi^*(x)\cdot \nabla
  \phi_v(x-y_j) \\ & \quad + \half \int_{\Bb_R} v(d(x,y_j)) \psi(x)^*
  \phi_v(x-y_j) \nu(x-y_j) \,.
\end{align}
By using the Cauchy-Schwarz inequality, we can obtain the upper bound
\begin{align}
   |A|^2 &\leq \int_{\Bb_R} dx \left[ |\nabla\xi(x)|^2 + \half
     v(d(x,y_j)) |\psi(x)|^2\right] \\ \nonumber & \quad \times
   \int_{\Bb_R}dx \left[ |\nabla \phi_v(x-y_j)|^2 + \half v(d(x,y_j))
     |\phi_v(x-y_j)|^2\right] |\nu(x-y_j)|^2\,.
\end{align}
For an upper bound, we can replace the integration region $\Bb_R$ in
the second integral by $\R^3$.  Since $\phi_v(x)$ is a radial
function, the angular integration in then can be performed by using
$\int_{\Ss^2} |\nu|^2 = 1$. The remaining expression is then bounded
by $a$ because of $\int_{\R^3} dx\, \left( |\nabla \phi_v(x)|^2 +
  \half v(|x|) |\phi_v(x)|^2 \right)= 4\pi a$. Hence we arrive at
\begin{equation}\label{comb}
\int_{\Bb_R} dx \left[ |\nabla\xi(x)|^2 
+ \half v(d(x,y_j)) |\psi(x)|^2\right] \geq \frac {|A|^2}a \,
\end{equation}
for any choice of $\nu$ as above. It remains to derive a lower bound
on $|A|^2$.

By partial integration, 
\begin{align}\nonumber
  &\int_{\Bb_R} dx\, \nu(x-y_j) \nabla\xi^*(x)\cdot \nabla
  \phi_v(x-y_j) \\ \nonumber &= - \int_{\Bb_R} dx\, \xi^*(x)
  \nu(x-y_j) \Delta \phi_v(x-y_j) \\ &\quad + \int_{\partial \Bb_R}
  d\omega_R\, \xi^*(x) \nu(x-y_j) n\cdot \nabla \phi_v(x-y_j) \,,
\end{align}
where $d\omega_R$ denotes the surface measure of the boundary of
$\Bb_R$, denoted by $\partial \Bb_R$, and $n$ is the outward normal
unit vector. Here we used the fact that $\nabla \nu(x)\cdot \nabla
\phi_v(x)=0$.  Now, by definition of $h(x)$, $\xi(x)= \psi(x) -
(2\pi)^{-3/2} h * \psi(x)$, where $*$ denotes convolution, i.e.,
$h*\psi(x) = \int_\Lambda dy\, h(x-y)\psi(y)$.  Using the zero-energy
scattering equation (\ref{scatteq}) for $\phi_v$, we thus see that
\begin{align}\nonumber
  A &= \int_{\partial \Bb_R} d\omega_R\, \psi^*(x) \nu(x-y_j) n\cdot
  \nabla \phi_v(x-y_j) \\ \nonumber &\quad - (2\pi)^{-3/2}
  \int_{\partial\Bb_R} d\omega_R\, (h*\psi)^*(x)
  \nu(x-y_j)n\cdot\nabla \phi_v(x-y_j) \\ &\quad + (2\pi)^{-3/2}
  \int_{\Bb_R}dx\, ( h*\psi)^*(x) \nu(x-y_j) \Delta \phi_v(x-y_j)\,.
\label{of}
\end{align} 
The last two terms on the right side of (\ref{of}) can be written as
\begin{equation}\label{expr}
(2\pi)^{-3/2} \int_\Lambda dx\, \psi^*(x) 
\left[ \int_{\Bb_R} d\mu(y)\,  h(y-x))\right]\,,
\end{equation}
where $d\mu$ is a (non-positive) measure supported in $\Bb_R$.
Explicitly, $d\mu(x) = \nu(x-y_j) \Delta \phi_v(x-y_j) dx - n\cdot
\nabla \phi_v(x-y_j)\nu(x-y_j) d\omega_R$, the second part being
supported on the boundary $\partial\Bb_R$.  Note that $\int_{\Bb_R}
d\mu = 0$, and also $\int_{\Bb_R} d|\mu| \leq 2 a \int_{\Ss^2} |\nu|
\leq 2 a \sqrt{4\pi}$ (by Schwarz's inequality).  Hence
\begin{equation}
  \left| \int_{\Bb_R} d\mu(y)\,  h(y-x)\right| \leq 2 a \sqrt{4\pi} f_R(x-y_j)\,,
\end{equation}
with $f_R$ defined in (\ref{deffr}). The expression (\ref{expr}) is
thus bounded from below by
\begin{align}\nonumber
   (\ref{expr}) & \geq - (2\pi)^{-3/2} 2a \sqrt{4\pi} \int_\Lambda dx\,
   |\psi(x)| f_R(x-y_j) \\ & \geq - a \left( \int_\Lambda dx\, |\psi(x)|^2
     w_R(x-y_j)\right)^{1/2}\,. \label{15}
\end{align}
Here, we used Schwarz's inequality as well as the definition of $w_R$
(\ref{defwr}) in the last step. Note that this last expression is
independent of $\nu$.

The only place where $\nu$ still enters is the first term on the right
side of (\ref{of}). By construction, $\nu$ depends only on the
direction of the line originating from $y_j$, which hits the boundary
of $\Bb_R$ at a distance not greater than $R$.  We distinguish two
cases. First, assume that the line hits the boundary at a distance
$R$. In this case, we choose $\nu$ to be equal to the value of $\psi$
at this boundary point. Secondly, if the length of the line is
strictly less than $R$, we then choose $\nu$ to be zero. Of course we
also have to normalize $\nu$ appropriately. The integrals are thus
only over the part of the boundary of $\Bb_j$ which is at a distance
$R$ from $y_j$. Let us denote this part of $\partial \Bb_R$ by
$\widetilde{\partial \Bb_R}$, assuming for the moment that it is not
empty.  We then have
\begin{equation}\label{thuso}
\int_{\partial \Bb_R} d\omega_R\, \psi^*(x) \nu(x-y_j) n\cdot \nabla
   \phi_v(x-y_j)
  = R \frac { \int_{\widetilde{\partial \Bb_R}} d\omega_R\,  |\psi(x)|^2
     n\cdot \nabla \phi_v(x-y_j)}{\left( \int_{\widetilde{\partial \Bb_R}}
       d\omega_R\, |\psi(x)|^2\right)^{1/2} } \,.
\end{equation}
We note that $n\cdot \nabla \phi_v(x-y_j) = a/R^2$ on
$\widetilde{\partial \Bb_R}$.  We thus obtain from
(\ref{of})--(\ref{thuso})
\begin{equation}
A \geq \frac aR\left( \int_{\widetilde{\partial \Bb_R}} d\omega_R\, |\psi(x)|^2
   \right)^{1/2} 
-  a \left( \int_\Lambda dx\, |\psi(x)|^2 w_R(x-y_j)\right)^{1/2}. 
\end{equation}
With the aid of Schwarz's inequality, we see that, for any $\eps>0$, 
\begin{equation}\label{comb2}
|A|^2 \geq \frac{a^2}{R^2} (1-\eps)\int_{\widetilde{\partial \Bb_R}}
d\omega_R\, |\psi(x)|^2  - \frac {a^2}\eps \int_\Lambda dx\, 
|\psi(x)|^2 w_R(x-y_j)\,.
\end{equation}
At this point we can also relax the condition that
$\widetilde{\partial \Bb_R}$ be non-empty; in case it is empty,
(\ref{comb2}) holds trivially.

In combination with (\ref{comb}), (\ref{comb2}) proves the desired
result (\ref{toshow}) in the special case when $U(|x|)$ is a radial
$\delta$-function sitting at a radius $R$, i.e., $U(|x|)= R^{-2}
\delta(|x|-R)$. The case of a general potential $U(|x|)$ follows
simply by integrating this result (i.e., Ineq. (\ref{toshow}) for this
special $U(|x|)$) against $U(R) R^2 dR$, noting that $\int dR\, U(R)
R^2\leq 1$ and that $w_R(x)$ is pointwise monotone increasing in $R$.


\begin{thebibliography}{99}
  
\bibitem{ber} F.A. Berezin, {\it Covariant and contravariant symbols
    of operators}, Izv. Akad. Nauk, Ser. Mat. {\bf 36}, 1134--1167
  (1972); English translation: USSR Izv. {\bf 6}, 1117--1151 (1973).
  F.A. Berezin, {\it General concept of quantization}, Commun. Math.
  Phys. {\bf 40}, 153--174 (1975).

\bibitem{dyson} F.J. Dyson, {\it Ground-State Energy of a Hard-Sphere Gas},
Phys. Rev. \textbf{106}, 20--26 (1957).

\bibitem{hsdec} C. Hainzl and R. Seiringer, {\it General Decomposition
    of Radial Functions on $\R^n$ and Applications to $N$-Body Quantum
    Systems}, Lett. Math.  Phys. {\bf 61}, 75--84 (2002).

\bibitem{huang}
K. Huang, {\it Statistical Mechanics}, $2^{\rm nd}$ ed., Wiley (1987).

\bibitem{lie}
E.H. Lieb, {\it The classical limit of quantum spin systems},
Commun. Math. Phys. {\bf 31}, 327--340 (1973).

\bibitem{LRuskai} E.H.\ Lieb, M.B.\ Ruskai, \textit{Proof of the
    strong subadditivity of quantum-mechanical entropy}, J. Math.
  Phys. {\bf 14}, 1938--1941 (1973). E.H.\ Lieb, M.B.\ Ruskai,
  \textit{A Fundamental Property of Quantum Mechanical Entropy}, Phys.
  Rev. Lett. {\bf 30}, 434--436 (1973).

\bibitem{LSSfermi} E.H. Lieb, R. Seiringer, and J.P. Solovej, {\it
     Ground State Energy of the Low Density Fermi Gas}, Phys. Rev. A
   {\bf 71}, 053605 (2005).
   
 \bibitem{oberw} E.H. Lieb, R. Seiringer, J.P. Solovej, and J.
   Yngvason, {\it The Mathematics of the Bose Gas and its
     Condensation}, Oberwolfach Seminars, Vol. 34, Birkh\"auser
   (2005).

\bibitem{LSY00} E.H. Lieb, R. Seiringer, and J. Yngvason, {\it Bosons
in a Trap: A Rigorous Derivation of the Gross-Pitaevskii Energy
Functional}, Phys. Rev. A {\bf 61}, 043602-1--13 (2000).

\bibitem{lsybog} E.H. Lieb, R. Seiringer, and J. Yngvason, {\it
     Justification of c-Number Substitutions in Bosonic Hamiltonians},
   Phys. Rev. Lett. {\bf 94}, 080401 (2005).

\bibitem{LY1998}
E.H.\ Lieb and J.\ Yngvason, {\it Ground State Energy of the Low
Density Bose Gas}, Phys. Rev. Lett. \textbf{80}, 2504--2507 (1998).

\bibitem{LY2d} E.H. Lieb and J. Yngvason, {\it The Ground State Energy
    of a Dilute Two-Dimensional Bose Gas}, J. Stat. Phys.
  \textbf{103}, 509--526 (2001).
   
 \bibitem{ohya} M. Ohya and D. Petz, {\it Quantum Entropy and Its
     Use}, Texts and Monographs in Physics, Springer (2004).

\bibitem{rob} D.W. Robinson, {\it The Thermodynamic Pressure in
     Quantum Statistical Mechanics}, Springer Lecture Notes in Physics,
   Vol. 9 (1971).

\bibitem{rue} D. Ruelle, {\it Statistical Mechanics. Rigorous
     Results}, World Scientific (1999).
   
 \bibitem{FermiT} R. Seiringer, {\it The Thermodynamic Pressure of a
     Dilute Fermi Gas}, Commun.  Math. Phys. {\bf 261}, 729--758
   (2006).

\bibitem{RSjellium} R. Seiringer, {\it A Correlation Estimate for
    Quantum Many-Body Systems at Positive Temperature}, Rev. Math.
  Phys. {\bf 18}, 233--253 (2006).

\end{thebibliography}
\end{document}